\documentclass[12pt]{iopart}
\usepackage{iopams}  
\usepackage{graphicx}
\begin{document}

\newcommand{\apj}{{Astrophys.\ J. }}
\newcommand{\apjs}{{Astrophys.\ J.\ Supp. }}
\newcommand{\apjl}{{Astrophys.\ J.\ Lett. }}
\newcommand{\aj}{{Astron.\ J. }}
\newcommand{\prl}{{Phys.\ Rev.\ Lett. }}
\newcommand{\prd}{{Phys.\ Rev.\ D }}
\newcommand{\mnras}{{Mon.\ Not.\ R.\ Astron.\ Soc. }}
\newcommand{\araa}{{ARA\&A }}
\newcommand{\nat}{{Nature }}
\newcommand{\cqg}{{Class.\ Quantum Grav.\ }}

\def\HS{{\mathfrak H}_3}
\def\is{\hbox{\scriptsize\rm i}}
\def\i{\hbox{\rm i}}
\def\j{\hbox{\rm j}}
\def\3{{\ss}}

\hfill astro-ph/0403597

\title[Horned Universes]
{Hyperbolic Universes with a Horned Topology and the CMB Anisotropy}

\author{Ralf Aurich$^1$, Sven Lustig$^1$, Frank Steiner$^1$ and Holger Then$^2$}

\address{$^1$Abteilung Theoretische Physik, Universit\"at Ulm,\\
Albert-Einstein-Allee 11, D-89069 Ulm, Germany}

\address{$^2$School of Mathematics, University of Bristol,\\
University Walk, Bristol, BS8~1TW, United Kingdom}

\begin{abstract}
We analyse the anisotropy of the cosmic microwave background (CMB)
in hyperbolic universes possessing a non-trivial topology
with a fundamental cell having an infinitely long horn.
The aim of this paper is twofold.
On the one hand, we show that the horned topology does not lead to
a flat spot in the CMB sky maps in the direction of the horn
as stated in the literature.
On the other, we demonstrate that a horned topology having a
finite volume could explain the suppression of the lower multipoles
in the CMB anisotropy as observed by COBE and WMAP.
\end{abstract}





\section{Introduction}

A fundamental problem in cosmology is the large-scale geometry of the
Universe, in particular its spatial curvature and topology.
Since the Einstein gravitational field equations are differential equations,
they determine the local properties of space-time, but not the global
structure of the Universe at large.
In the so-called concordance model of cosmology
it is assumed that the Universe is at large scales spatially flat
and possesses the trivial topology, implying that it has infinite volume.
In the framework of inflationary scenarios, these properties are supposed
to be determined by the initial conditions at the Big Bang.
However, since we are lacking a theory of quantum gravity,
the initial conditions cannot be derived from first principles.
Instead, one can analyse the recent astronomical data, in particular
on the cosmic microwave background (CMB) radiation, in order to
deduce the curvature and the topology of the Universe at large scales.

In 1992, COBE \cite{Smoot_et_al_1992} made the spectacular discovery of
the temperature fluctuations of the CMB and thus provided important clues
about the early Universe and its time evolution.
Expanding the observed temperature fluctuations $\delta T(\hat n)$
across the microwave sky into spherical harmonics $Y_l^m(\hat n)$,
yields the expansion coefficients $a_{lm}$ which in turn lead to
the {\it multipole moments}
\begin{equation}
\label{Eq:C_l}
C_l \; := \;
\frac 1{2l+1} \, \sum_{m=-l}^l \, | a_{lm} |^2
\end{equation}
and the {\it angular power spectrum}
$\delta T_l^2 := l(l+1) C_l /(2\pi)$.
In particular, COBE \cite{Smoot_et_al_1992} detected in the
angular power spectrum of the CMB a low quadrupole moment $C_2$
corresponding to a strange suppression of power on large angular scales.
It was soon realized that although the standard
cosmological models are in agreement with the CMB anisotropy on small and
medium scales, they fail to match the loss of power on large angular scales,
especially those corresponding to the quadrupole moment.
This observation was one of the motivations to study non-trivial topologies
(see the reviews \cite{Lachieze-Rey_Luminet_1995,Levin_2002}).
In particular, compact hyperbolic universes were studied by several authors
\cite{Bond_Pogosyan_Souradeep_1998,Bond_Pogosyan_Souradeep_1999a,%
Bond_Pogosyan_Souradeep_1999b,Cornish_Spergel_Starkman_1998,Aurich_1999,%
Cornish_Spergel_1999,Inoue_Tomita_Sugiyama_1999,Aurich_Steiner_2000}.
For these compact hyperbolic universes, it was shown
that the non-trivial topology leads indeed to a suppression of $C_l$
for small values of $l$.
Thus the observed loss of power on large angular scales was interpreted
as a clear hint to a non-trivial topology of our Universe.

\begin{figure}[htb]
\begin{center}
\hspace*{-12pt}\begin{minipage}{9cm}
\includegraphics[width=9cm]{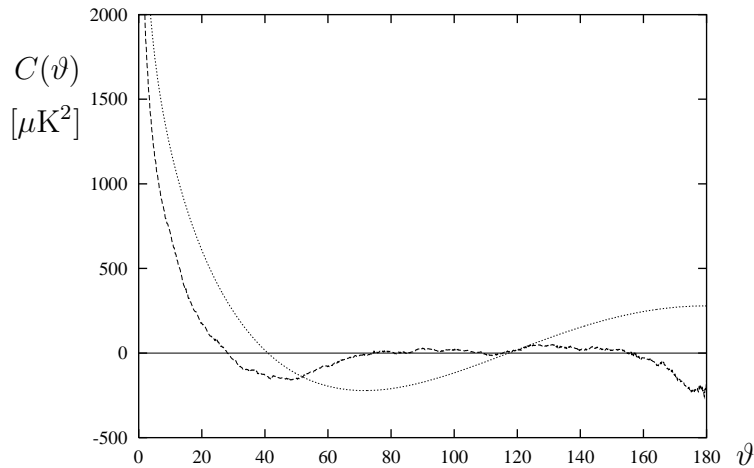}
\put(-1,4){$\vartheta$}
\put(-274,150){$C(\vartheta)$}
\put(-276,130){$[\mu \hbox{K}^2]$}
\end{minipage}
\vspace*{-10pt}
\end{center}
\caption{\label{Fig:C_theta_concordance}
The correlation function $C(\vartheta)$ from the WMAP-data (dashed curve)
in comparison with the concordance model (dotted curve).
}
\end{figure}

The first findings of NASA's explorer mission
``Wilkinson Microwave Anisotropy Probe'' (WMAP) \cite{Hinshaw_et_al_2003}
have tremendously increased our knowledge of the temperature fluctuations
of the CMB, since WMAP has measured the anisotropy  of the CMB radiation
over the full sky with high accuracy.
The WMAP data confirm  not only COBE's measurement of the low quadrupole
moment $C_2$, but display in the temperature (auto-) correlation function
$C(\vartheta)$ $\,(\hat n \cdot \hat n' = \cos \vartheta)$
\begin{equation}
\label{Eq:C_theta}
C(\vartheta) \; \simeq \;
\frac 1{4\pi} \, \sum_{l=2}^\infty \, (2l+1) \, C_l \, P_l(\cos\vartheta)
\end{equation}
very weak correlations at wide angles,
$70^\circ \lesssim \vartheta \lesssim 150^\circ$,
see the dashed curve in figure \ref{Fig:C_theta_concordance}.
(Note that the monopole and dipole are not included in the sum
(\ref{Eq:C_theta}).)
At the largest angles, $\vartheta \gtrsim 160^\circ$, the WMAP-data
display even a ``correlation hole'',
i.\,e.\ negative values of $C(\vartheta)$.
In figure \ref{Fig:C_theta_concordance} we also show as a dotted curve
the theoretical prediction according to the concordance model using the
best-fit values for the cosmological parameters as obtained by WMAP
\cite{Hinshaw_et_al_2003}.
It is seen that the concordance model,
i.\,e.\ the best-fit $\Lambda$CDM model,
does not reproduce the experimentally observed suppression at
$\vartheta \gtrsim 60^\circ$ and the observed correlation hole.

Since the correlation function $C(\vartheta)$ emphasizes large angular scales
and thus the low $l$-range, it is an ideal indicator function
to search for a fingerprint of a possible non-trivial topology of
the Universe.

The evidence for a low quadrupole and missing power on large scales
is, however, currently heavily discussed.
The low value of the quadrupole obtained by the WMAP team
\cite{Hinshaw_et_al_2003} has been criticized by the way
how the foregrounds \cite{Bennett_et_al_2003b} are taken into account
which are mainly caused by free-free, synchrotron and dust emission.
Alternative reconstructions of the true cosmological signal from the
five frequency bands measured by WMAP are discussed in
\cite{Tegmark_deOliveira_Costa_Hamilton_2003,Eriksen_Banday_Gorski_Lilje_2004}
which lead to a higher quadrupole moment.
The observation that the plane of the quadrupole and two of the
three planes of the octopoles are aligned towards the ecliptic might be seen
as a hint of an unknown source or sink of CMB radiation in the
outer solar system or as an unrecognized systematic
\cite{Schwarz_Starkman_Huterer_Copi_2004}.
The influence of the approximation of the likelihood function used
for the angular power spectra analysis is investigated in
\cite{Slosar_Seljak_Makarov_2004}, where for the lowest multipoles
an exact likelihood function estimation is presented.
The statistical significance of the low value of the quadrupole moment
is also discussed in \cite{Efstathiou_2003} who found the
WMAP results to be consistent with the concordance $\Lambda$CDM model.
The statistic is limited by the one-sky realization we are able to measure,
also called cosmic variance.
To circumvent this problem it is suggested
to utilize the fact that the CMB polarization is sourced
by the local temperature quadrupole.
On the one hand, the polarization signal contains information about the
quadrupole moment at the reionization epoch.
This could lead to a probability for the observed quadrupole moment
of the order of $10^{-4}$ \cite{Skordis_Silk_2004}
compared with the concordance model.
On the other hand, future measurements of the linear polarization
of the CMB towards clusters of galaxies can betray the local
quadrupole moment at the location of the cluster
\cite{Kamionkowski_Loeb_1997,Sazonov_Sunyaev_1999,Seto_Sasaki_2000}
thus circumventing the cosmic variance limit.

Furthermore, we would like to mention that ``the question
whether the geometry of the three-dimensional space of astronomy
might be non-Euclidean'' \cite{Chandrasekhar_1986}
has already been posed by Schwarzschild
\cite{Schwarzschild_1992,Schwarzschild_1900,Schwarzschild_1998}
in 1900, fifteen years before the founding of general relativity!
He stated the problem as follows (see p.\,32 in \cite{Schwarzschild_1992}):
``As must be known to you, during this century [meaning the 19th century]
one has developed non-Euclidean geometry (besides Euclidean geometry),
the chief examples of which are the so-called spherical and
pseudo-spherical spaces.
We can wonder how the world would appear in a spherical or
pseudo-spherical geometry with possibly a finite radius of curvature \dots
One would then find oneself, if one will, in a geometrical fairyland;
and one does not know whether the beauty of this fairyland may in fact
be realized in nature.''
Schwarzschild ``actually estimated limits to the radius of curvature
of the three-dimensional space with the astronomical data available
at his time and concluded that if the space is hyperbolic
its radius of curvature cannot be less than 64 light years and
that if the space is spherical its radius of curvature must
at least be 1600 light years'' \cite{Chandrasekhar_1986}.
In an Addendum to his article
``On the permissible scale of curvature of space'' \cite{Schwarzschild_1900}
he mentioned also the possibility of spaces with non-trivial topology
by referring to Clifford-Klein space forms.
And he emphasized that such spaces do not necessarily lead to infinite
universes as commonly assumed, even in the case of Euclidean or
hyperbolic geometry.
Schwarzschild concluded that the only condition imposed by astronomical
data is that the volume of the Universe must be larger than the visible
system of stars.

It is the purpose of this paper to investigate two models of the Universe
whose global topology is not the universal covering space of
their spatial geometry.
The spatial geometry of the models is the pseudo-spherical or
hyperbolic space, i.\,e.\ it has negative curvature.
The corresponding hyperbolic universes are non-compact and possess an
infinitely long horn.
The first model was introduced in 1976 by Sokolov and Starobinskii
\cite{Sokolov_Starobinskii_1976} and is described by a fundamental
cell ${\cal F}$ having infinite volume.
The second model, called the Picard model \cite{Picard_1884},
has also an infinitely long horn,
but nevertheless possesses a finite volume.
The Picard model has been investigated in detail in the context
of quantum chaos \cite{Matthies_1995,Steil_1999,Then_2003}.

The Sokolov-Starobinskii model has been studied already in
\cite{Sokolov_Starobinskii_1976} and \cite{Levin_Barrow_Bunn_Silk_1997}
(see also the review \cite{Levin_2002}).
In \cite{Levin_Barrow_Bunn_Silk_1997} it has been claimed that the
periodic horn topology produces a flat spot in the temperature map
of the CMB even when multiple images of astronomical sources are
unobservable.
The flat spot, i.\,e.\ a large flat region in the CMB sky map
corresponding to a negligibly small metric perturbation, is argued
in \cite{Levin_Barrow_Bunn_Silk_1997} to result from the
exponential decay of the eigenmodes in the direction of the horn.

We shall show in the following that the flat spot found in
\cite{Levin_Barrow_Bunn_Silk_1997} is a consequence of a too low
wavenumber cut-off $k_c$.
For the cut-off chosen in \cite{Levin_Barrow_Bunn_Silk_1997},
the eigenmodes cannot produce indeed a perturbation in the
horn at the position which corresponds to the distance of the
surface of last scattering (SLS).
As a consequence, the sky maps computed in \cite{Levin_Barrow_Bunn_Silk_1997}
do not show temperature fluctuations in the horn.
However, if the cut-off is sufficiently increased, we shall demonstrate
that there is no suppression  in the horn and therefore no flat spot.

The universes described by the Sokolov-Starobinskii and Picard model,
respectively, possess negative spatial curvature in contrast
to the concordance model corresponding to a spatially flat universe.
Therefore the question arises whether a negatively curved universe
is realized in nature.
In \cite{Aurich_Steiner_2002b,Aurich_Steiner_2003} we have analysed
the CMB data and the magnitude redshift relation of supernovae type Ia
in the framework of quintessence models
and have shown that these data are consistent with a nearly flat
hyperbolic geometry of our Universe
if the optical depth $\tau$ to the SLS is not too big.
The restriction comes from the large amplitude of the fluctuations at
large scales in the CMB.
However, it was shown in \cite{Conversi_Melchiorri_Mersini_Silk_2004}
that by replacing the quintessence component having a rest frame sound
velocity of $c_s = c$ by a generalized dark matter component
\cite{Hu_1998} with a vanishing rest frame sound velocity $c_s = 0$,
the amplitude in the CMB anisotropy is much smaller at large scales.
Thus with such a generalized dark matter component universes with
negative curvature are permissible even for larger optical depths $\tau$.

Recently, an ellipticity analysis of the CMB maps has been reported
\cite{Gurzadyan_et_al_2003a,Gurzadyan_et_al_2003b,Gurzadyan_et_al_2004}
which gives further support to a hyperbolic spatial geometry of the
Universe.
In these analyses hot and cold anisotropy spots in the CMB maps have been
studied in terms of shape for various temperature thresholds.
Analysing with the same algorithm the COBE-DMR, BOOMERanG 150 GHz
and WMAP maps, an ellipticity of the anisotropy spots has been found of
the same average value (around 2) for these experiments.
The WMAP data confirm the effect for scales both smaller and larger
than the horizon at the SLS.
This suggests that the effect is not due to physical effects at the SLS,
and can arise after, while the photons are moving freely in the Universe.

Finally, we would like to mention that recently a model was presented
\cite{Luminet_Weeks_Riazuelo_Lehoucq_Uzan_2003} possessing positive
curvature and a finite volume with the shape of the
Poincar\'e dodecahedral space.
The authors of ref.\ \cite{Luminet_Weeks_Riazuelo_Lehoucq_Uzan_2003}
calculated the CMB multipoles for $l=2,3$ and 4, set the overall
normalization factor to match the WMAP data at $l=4$ and examined the
prediction for $l=2$ and 3.
They found a weak suppression of the power at $l=3$ and strong
suppression at $l=2$ in agreement with the WMAP observations.
However, in ref.\ \cite{Luminet_Weeks_Riazuelo_Lehoucq_Uzan_2003}
only the modes up to $k_c=30$ have been used, and it thus remains
the question about how this low cut-off affects the integrated
Sachs-Wolfe (ISW) contribution.
Our experience shows that increasing the cut-off usually enhances
the ISW contribution.


\section{The CMB anisotropy}

The CMB anisotropy is computed for universes whose background models are
described by the Friedmann equation $(c=1)$
\begin{equation}
\label{Eq:Friedmann}
H^2 := \left(\frac{a'}{a^2}\right)^2 \; = \;
\frac{8\pi G}3 \varepsilon_{\hbox{\scriptsize tot}} - \frac K{a^2}
\hspace{10pt} , \hspace{10pt}
K \in \{ -1,0,+1\}
\hspace{10pt} .
\end{equation}
Here $a=a(\eta)$ denotes the cosmic scale factor as a function
of conformal time $\eta = \int \frac{dt}a$ and $K$ the spatial curvature.
The total energy density $\varepsilon_{\hbox{\scriptsize tot}}$
is given by a sum of 4 components: radiation, baryonic matter,
cold dark matter and a positive cosmological constant $\Lambda$.
In this paper, we consider only scalar perturbations,
i.\,e.\ vector and tensor modes are neglected.
Furthermore, we are mainly interested in fluctuations on scales
larger than the size of the horizon at the time of recombination
since these fluctuations should contain the fingerprint of a
non-trivial topology.
Above the size of the horizon the temperature anisotropy can be computed
neglecting physical processes important only on small scales
such as the Silk damping.
The metric with scalar perturbations is written
in the conformal-Newtonian gauge in terms of scalar functions $\Phi$ and
$\Psi$ as
$$
ds^2 \; = \; a^2(\eta) \left\{ (1+2\Phi) d\eta^2 - (1-2\Psi)
\gamma_{ij} dx^i dx^j \right\}
\hspace{10pt} ,
$$
where $\Phi=\Psi$ for an energy-momentum tensor $T_{\mu\nu}$ with
$T_{ij}=0$ for $i\neq j$ and $i,j=1,2,3$.
($\gamma_{ij}$ denotes the metric of the 3-space.)
The metric perturbation $\Phi$ is expanded into the eigenfunctions
$\psi_k(\vec x\,)$ of the (negative) Laplace-Beltrami operator $\Delta$,
i.\,e.\ $(\Delta+E)\psi_k(\vec x\,) = 0$ with $k = \sqrt{E+K}$,
e.\,g.\ for a discrete spectrum
\begin{equation}
\label{Eq:Metric_expansion}
\Phi(\eta,\vec x\,) \; = \;
\sum_{n=1}^\infty \Phi_{k_n}(\eta) \; \psi_{k_n}(\vec x\,)
\hspace{10pt} .
\end{equation}

In the {\it single fluid approximation}, the radiation, the baryonic matter
and the cold dark matter are treated as one fluid which is valid
on scales larger than the horizon size at recombination.
There the entropy perturbation $\delta S$ is thus negligible,
i.\,e.\ $\delta S=0$.
Then the evolution of the metric perturbation $\Phi(\eta,\vec x\,)$
gives in first-order perturbation theory in the
conformal-Newtonian gauge \cite{Mukhanov_Feldman_Brandenberger_1992}
\begin{equation}
\label{Eq:Phi_first-order}
\Phi'' + 3 \hat H (1+c_s^2) \Phi' \, - \, c_s^2 \Delta \Phi \, + \,
\{2 \hat H' + (1+3c_s^2)(\hat H^2-K)\} \Phi \; = \; 0
\hspace{10pt} ,
\end{equation}
where $\hat H := a'/a$.
The quantity
$c_s^2 = \left(3 + \frac 94 \varepsilon_{\hbox{\scriptsize mat}}/
\varepsilon_{\hbox{\scriptsize r}}\right)^{-1}$
can be interpreted as the sound velocity.
Specifying the initial perturbation $\Phi$ at $\eta=0$ such that it
corresponds to a scale-invariant Harrison-Zel'dovich spectrum
(see \cite{Aurich_Steiner_2000}; $\hat \eta = 
2 \sqrt{\Omega_{\hbox{\scriptsize r}}\Omega_{\hbox{\scriptsize curv}}}
/\Omega_{\hbox{\scriptsize mat}}$)
\begin{equation}
\label{Eq:Harrison_Zeldovich}
\Phi_{\vec k}(0) \; = \; \frac{\alpha}{\sqrt{k(k^2-K)}}
\hspace{10pt} \hbox{ and } \hspace{10pt}
\Phi_{\vec k}'(0) \; = \; - \, \frac{\Phi_{\vec k}(0)}{8\hat \eta}
\hspace{10pt} ,
\end{equation}
where $\Phi_{\vec k}(\eta)$ denote the expansion coefficients
of $\Phi(\eta,\vec x\,)$ (see equations (\ref{Eq:Metric_expansion_SoSta_cusp})
and (\ref{Eq:Metric_expansion_SoSta_sin_cos}) below),
allows the computation of the time-evolution of the metric
perturbation $\Phi$.
Here we have defined the density parameters
$\Omega_x := \varepsilon_x(\eta_0)/\varepsilon_{\hbox{\scriptsize crit}}$
with $\varepsilon_{\hbox{\scriptsize crit}}=3 H_0^2/(8\pi G)$
for $x=\hbox{r (=radiation)}$,
mat (= baryonic + cold dark matter (CDM)), and
curv (= curvature) with
$\Omega_{\hbox{\scriptsize curv}} = 1 - \Omega_{\hbox{\scriptsize tot}}$.
The Hubble constant $H_0 = h \times 100 \hbox{ km s}^{-1} \hbox{Mpc}^{-1}$
is set in the following to $h=0.65$.
($\alpha$ is a normalization constant and $\eta_0$ denotes the
conformal time at the present epoch.)
The metric perturbation $\Phi$ in turn gives the input to the
Sachs-Wolfe formula \cite{Sachs_Wolfe_1967}
which reads for isentropic initial conditions
\begin{equation}
\label{Eq:Sachs_Wolfe}
\frac{\delta T}T(\hat n) \; = \;
2 \Phi(\eta_{\hbox{\scriptsize{SLS}}},\vec x(\eta_{\hbox{\scriptsize{SLS}}}))
- \, \frac 32 \Phi(0,\vec x(0)) +
2 \, \int_{\eta_{\hbox{\scriptsize{SLS}}}}^{\eta_0} d\eta
\frac{\partial\Phi(\eta,\vec x(\eta))}{\partial\eta}
\hspace{5pt} ,
\end{equation}
from which one obtains the desired temperature fluctuations $\delta T$
of the CMB.
Here $\hat n$ denotes the unit vector in the direction of
$\vec x(\eta_{\hbox{\scriptsize{SLS}}})-\vec x(\eta_0)$,
i.\,e.\ in the direction from which the photons arrive.
The single fluid approximation has been used in a lot of earlier papers
concerning the CMB anisotropy for non-trivial topologies.
We summarized this approach here, because we need it for the comparison
with earlier papers.

We use in our computations the {\it tight-coupling approximation}
in which only the radiation and the baryonic matter is approximated
as a single fluid before recombination,
whereas the cold dark matter component is independent.
The following summary of the tight-coupling approximation follows
the lines of Mukhanov \cite{Mukhanov_2003}.
The Sachs-Wolfe formula reads for the tight-coupling approximation
\begin{eqnarray} \nonumber
\frac{\delta T}T(\hat n) & = &
\int d^3k \; \left[ \left( \Phi_{\vec k}(\eta) +
\frac{\delta_{\vec k,\hbox{\scriptsize{r}}}(\eta)}4 +
\frac{a(\eta) V_{\vec k,\hbox{\scriptsize{r}}}(\eta)}{E_{\vec k}}
\frac{\partial}{\partial \tau} \right) \psi_{\vec k}(\tau(\eta),\theta,\phi)
\right]_{\eta=\eta_{\hbox{\scriptsize{SLS}}}}
\\ & & \hspace{10pt}
\label{Eq:Sachs_Wolfe_tight_coupling}
\, + \,
2 \int d^3k \; \int_{\eta_{\hbox{\scriptsize{SLS}}}}^{\eta_0} d\eta \,
\frac{\partial\Phi_{\vec k}(\eta)}{\partial\eta} \,
\psi_{\vec k}(\tau(\eta),\theta,\phi)
\hspace{10pt} ,
\end{eqnarray}
where the $\vec k$-integration has to be interpreted as a summation in the
case of a discrete spectrum.
$(\tau(\eta),\theta,\phi)$ denote the spherical coordinates of the photon
path in the direction $\hat n$.
Here, $\delta_{\vec k,\hbox{\scriptsize{r}}}(\eta)$ is the expansion
coefficient of the relative perturbation in the radiation component, and
$V_{\vec k,\hbox{\scriptsize{r}}}(\eta)$ is the expansion
coefficient of the spatial covariant divergence
of the velocity field of the tightly coupled radiation-baryon components.
The term in equation (\ref{Eq:Sachs_Wolfe_tight_coupling})
which has to be evaluated at $\eta=\eta_{\hbox{\scriptsize{SLS}}}$
is computed using the tight-coupling approximation.
The second term, i.\,e.\ the integrated Sachs-Wolfe effect,
is computed with a decoupled baryon component.
In the case of isentropic initial conditions, the relation between the
perturbation in the radiation and baryonic component
during the tight-coupling phase is
$\delta_{\vec k,\hbox{\scriptsize{b}}} =
\frac 34\delta_{\vec k,\hbox{\scriptsize{r}}}$.
The time-evolution of the perturbations is determined by the following
system of differential equations which has to be solved
numerically.
For the metric perturbation, one obtains ($E_{\vec k} = k^2 - K$)
$$
\frac{d}{d \eta} \Phi_{\vec k}(\eta) \, + \,
\frac{a(\eta)}{a'(\eta)} \left( \frac 13 E_{\vec k}  +
\left(\frac{a'(\eta)}{a(\eta)}\right)^2 - K\right) \, \Phi_{\vec k}(\eta)
\; = \;
- \, \frac{4\pi G}3 \, \frac{a^3(\eta)}{a'(\eta)} \,
\delta\varepsilon_{\vec k}(\eta)
\hspace{10pt} ,
$$
where the energy perturbation $\delta\varepsilon_{\vec k}(\eta)$
before recombination is given by
$$
\delta\varepsilon_{\vec k}(\eta) \; = \;
\epsilon_{\hbox{\scriptsize{CDM}}}(\eta)
\delta_{\vec k,\hbox{\scriptsize{CDM}}}(\eta) \, + \,
\frac{\epsilon_{\hbox{\scriptsize{r}}}(\eta)}{3c_s^2}
\delta_{\vec k,\hbox{\scriptsize{r}}}(\eta)
\hspace{10pt} ,
$$
with
$c_s^2 = \left(3 + \frac 94 \varepsilon_{\hbox{\scriptsize b}}/
\varepsilon_{\hbox{\scriptsize r}}\right)^{-1}$,
and after recombination by
$$
\delta\varepsilon_{\vec k}(\eta) \; = \;
\epsilon_{\hbox{\scriptsize{CDM}}}(\eta)
\delta_{\vec k,\hbox{\scriptsize{CDM}}}(\eta) \, + \,
\epsilon_{\hbox{\scriptsize{b}}}(\eta)
\delta_{\vec k,\hbox{\scriptsize{b}}}(\eta) \, + \,
\epsilon_{\hbox{\scriptsize{r}}}(\eta)
\delta_{\vec k,\hbox{\scriptsize{r}}}(\eta)
\hspace{10pt} .
$$
The energy and momentum conservation equations for the CDM component read
$$
\frac{d}{d \eta} \left(
\delta_{\vec k,\hbox{\scriptsize{CDM}}}(\eta) - 3 \Phi_{\vec k}(\eta) \right)
\; = \; - \, a(\eta) V_{\vec k,\hbox{\scriptsize{CDM}}}(\eta)
$$
and
$$
\frac{d}{d \eta}
\left( a^2(\eta) V_{\vec k,\hbox{\scriptsize{CDM}}}(\eta) \right)
\; = \;
a(\eta) \, E_{\vec k} \, \Phi_{\vec k}(\eta)
\hspace{10pt} ,
$$
respectively.
For the radiation component, the energy conservation equation is
$$
\frac{d}{d \eta} \left(
\delta_{\vec k,\hbox{\scriptsize{r}}}(\eta) - 4 \Phi_{\vec k}(\eta) \right)
\; = \; - \, \frac 43 a(\eta) V_{\vec k,\hbox{\scriptsize{r}}}(\eta)
\hspace{10pt} ,
$$
whereas the momentum conservation equation is before recombination
given by
$$
\frac{d}{d \eta}
\left( \frac 1{c_s^2} \, a(\eta) V_{\vec k,\hbox{\scriptsize{r}}}(\eta) \right)
\; = \;
E_{\vec k} \, \left( \frac 1{c_s^2} \, \Phi_{\vec k}(\eta) \, + \,
\frac 34 \, \delta_{\vec k,\hbox{\scriptsize{r}}}(\eta) \right)
$$
and after recombination by
$$
\frac{d}{d \eta}
\left( a(\eta) V_{\vec k,\hbox{\scriptsize{r}}}(\eta) \right)
\; = \;
E_{\vec k} \left( \Phi_{\vec k}(\eta) +
\frac 14 \delta_{\vec k,\hbox{\scriptsize{r}}}(\eta) \right)
\hspace{10pt} .
$$
After recombination the energy and momentum conservation equations
for the baryonic component are analogous to that of the CDM component.
The initial conditions of the various components depend all on
that of $\Phi_{\vec k}(0)$ as
$$
\delta_{\vec k,\hbox{\scriptsize{r}}}(0) \; = \;
\frac 43 \, \delta_{\vec k,\hbox{\scriptsize{b}}}(0) \; = \;
\frac 43 \, \delta_{\vec k,\hbox{\scriptsize{CDM}}}(0) \; = \;
-2 \, \Phi_{\vec k}(0)
$$
and
$$
V_{\vec k,\hbox{\scriptsize{r}}}(0) \; = \;
V_{\vec k,\hbox{\scriptsize{b}}}(0) \; = \;
V_{\vec k,\hbox{\scriptsize{CDM}}}(0) \; = \;
E_{\vec k} \, \frac{H_0 |\Omega_{\hbox{\scriptsize{curv}}}|}
{2\sqrt{\Omega_{\hbox{\scriptsize{r}}}}} \, \Phi_{\vec k}(0)
\hspace{10pt} .
$$
The initial power spectrum $\Phi_{\vec k}(0)$ is taken as the usual
Harrison-Zel'dovich spectrum having a spectral index $n_S = 1$
which is in agreement with the current observations
\cite{Spergel_et_al_2003,Bridle_Lewis_Weller_Efstathiou_2003}.
Furthermore, we do not take a running spectral index into account since,
as shown in \cite{Bridle_Lewis_Weller_Efstathiou_2003},
only the first three multipoles $C_2$, $C_3$ and $C_4$ are responsible
for a non-vanishing running spectral index, and
it is just these multipoles which are most strongly modified by
a fundamental cell of finite volume.


\section{The Sokolov-Starobinskii model}

The model was introduced by Sokolov and Starobinskii
\cite{Sokolov_Starobinskii_1976} for cosmological studies.
In order to describe the model, we have to introduce the
hyperbolic three-space for which a wealth of models exists,
see e.\,g.\ \cite{Ratcliffe_1994}.
Here we choose the upper half-space model
\begin{equation}
\HS \; = \; \{ (x_1,x_2,x_3)\in{\mathbb R}^3 \; |\; x_3>0 \}
\hspace{10pt} ,
\end{equation}
equipped with the Riemannian metric
\begin{equation}
\label{Eq:H_metric}
ds^2 \; = \;\frac{dx_1^2+dx_2^2+dx_3^2}{x_3^2}
\end{equation}
corresponding to a constant Gaussian curvature $K=-1$.
The geodesics of a particle moving freely in the upper half-space
are straight lines and semicircles perpendicular to the
$x_1$-$x_2$-plane.

In the Sokolov-Starobinskii model \cite{Sokolov_Starobinskii_1976}
one identifies points periodically along $x_1$ and $x_2$ according to
$(x_1,x_2,x_3) \equiv (x_1+\mu a,x_2+\nu b,x_3)$
with $\mu,\nu\in \mathbb{Z}$.
The positive constants $a$ and $b$ define the toroidal topology.
The Sokolov-Starobinskii model has an infinitely long horn in the
positive $x_3$ direction
whose physical cross section decreases according to (\ref{Eq:H_metric})
with increasing $x_3$.
On the other hand, towards $x_3\to 0^+$ the cross section increases
without bound.
Thus the fundamental cell
${\cal F} =
\{ \vec x \in \mathbb{R}^3| 0 \leq x_1 < a, 0 \leq x_2 < b, x_3>0\}$
has infinite volume.

\begin{figure}[htb]
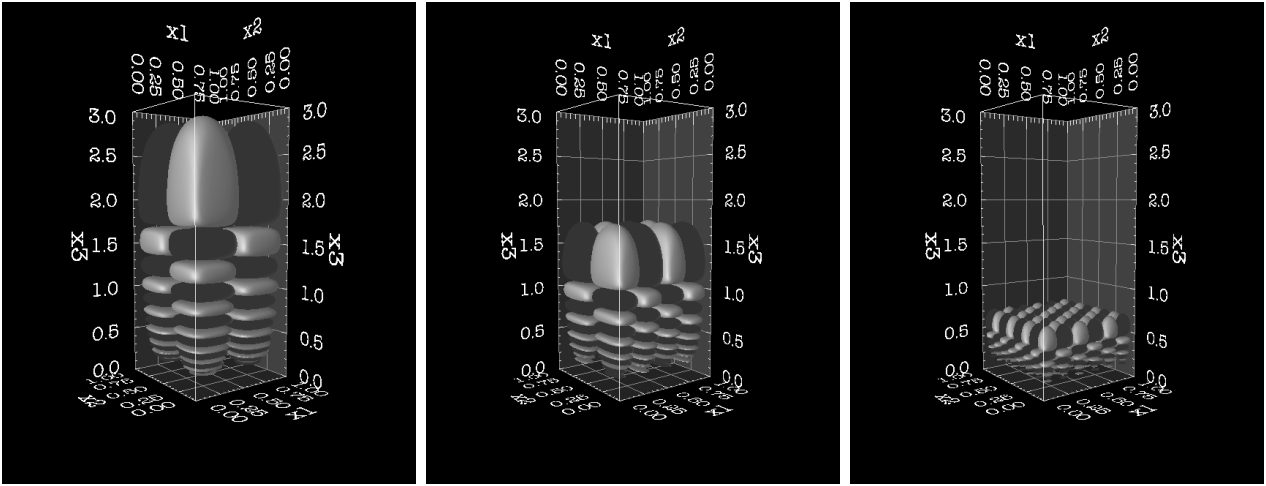

\begin{center}
\hspace*{-12pt}\begin{minipage}{18cm}
\includegraphics[width=5.5cm]{psplots/wfk_k_20_m_1_n_1_bw.epsf}
\includegraphics[width=5.5cm]{psplots/wfk_k_20_m_2_n_1_bw.epsf}
\includegraphics[width=5.5cm]{psplots/wfk_k_20_m_3_n_4_bw.epsf}
\end{minipage}
\vspace*{-10pt}
\end{center}
\caption{\label{Fig:decaying_modes}
Three eigenmodes (\protect\ref{Eq:SoSta_cusp_eigenmodes})
using two sines for the eigenvalue $k=20$ and $a=b=1$ are shown.
The eigenmodes belong to $(m,n)=(1,1)$, $(m,n)=(2,1)$ and
$(m,n)=(3,4)$, respectively, from left to right.
With increasing values of $Q$, the exponential suppression of the
$K$-Bessel function occurs earlier in the horn.
The suppression towards $x_3\to 0$ is due to the $x_3$-factor
in eq.\,(\protect\ref{Eq:SoSta_cusp_eigenmodes}).
}
\end{figure}

The eigenmodes of the Laplace-Beltrami operator on $\HS$
$$
\Delta \; = \; x_3^2 \, \left(
\frac{\partial^2}{\partial x_1^2} \, + \,
\frac{\partial^2}{\partial x_2^2} \, + \,
\frac{\partial^2}{\partial x_3^2} \right) \, - \,
x_3\, \frac{\partial}{\partial x_3}
$$
consist of decaying modes and plane waves.
The decaying modes are given by
$(k>0, (m,n) \in (\mathbb{N}_0 \times \mathbb{N}_0) - \{0,0\})$
\begin{equation}
\label{Eq:SoSta_cusp_eigenmodes}
\psi_{kmn}(\vec x\,) \; = \;
N_{kmn} \, x_3 \, K_{\is k}(Q x_3) \,
\left( \begin{tabular}[c]{c} $\sin \frac{2\pi m}a x_1$ \\
$\cos \frac{2\pi m}a x_1$ \end{tabular} \right) \,
\left( \begin{tabular}[c]{c} $\sin \frac{2\pi n}b x_2$ \\
$\cos \frac{2\pi n}b x_2$ \end{tabular} \right)
\hspace{10pt} ,
\end{equation}
with
$$
N_{kmn} \; = \;
\sqrt{\frac{2k \sinh \pi k}{\pi^2} \cdot
\frac{(2-\delta_{m0})(2-\delta_{n0})}{ab}}
\hspace{10pt} \hbox{ and } \hspace{10pt}
Q \; = \; 2\pi \sqrt{\left(\frac ma\right)^2 + \left(\frac nb\right)^2}
\; > \; 0
\hspace{10pt} .
$$
Here $K_{\is k}(z)$ denotes the modified Bessel function of order $\i k$.
Due to the above normalization factor $N_{kmn}$,
the eigenmodes (\ref{Eq:SoSta_cusp_eigenmodes}) are normalized as follows
$$
\int_{\cal F} d\mu \,
\psi_{kmn}(\vec x\,) \, \psi_{k'm'n'}(\vec x\,) \; = \;
\delta(k-k') \delta_{m,m'} \delta_{n,n'}
\hspace{5pt} \hbox{ with } \hspace{5pt}
d\mu = \frac{dx_1 dx_2 dx_3}{x_3^3}
\hspace{10pt} .
$$
Three examples of the eigenmodes (\ref{Eq:SoSta_cusp_eigenmodes})
for $k=20$ and three different pairs of $m$ and $n$
are shown in figure \ref{Fig:decaying_modes}.
The domain, where the modes dominate, decreases in $x_3$
with increasing $Q$.

The metric perturbation corresponding to these modes can then be expanded as
\begin{equation}
\label{Eq:Metric_expansion_SoSta_cusp}
\Phi(\eta,\vec x\,) \; = \;
\int_0^\infty dk \; \Phi_k(\eta) \; {\sum_{m,n}} ' \; r(k,m,n) \;
\psi_{kmn}(\vec x\,)
\hspace{10pt} ,
\end{equation}
where the prime at the sum indicates that the term with $m=n=0$ is excluded.
The $r(k,m,n)$ are real Gaussian expansion coefficients 
with mean value zero and variance independent of $k$
such that the amplitude $\Phi_k(\eta)$ describes the usual power spectrum
(\ref{Eq:Harrison_Zeldovich}).
This leads to the usual statistical properties of the fluctuations
far away from the horn.
The coefficients $r(k,m,n)$ do not carry any information about
the direction of the horn.
The information about the horn is only contained in the eigenmodes
of the Laplacian.

The eigenmodes (\ref{Eq:SoSta_cusp_eigenmodes})
decay exponentially in the horn due to the $K$-Bessel function.
In addition to (\ref{Eq:SoSta_cusp_eigenmodes}) there are further
eigenmodes which constitute non-decaying plane wave solutions in the horn.
They are given by $(k>0)$
\begin{equation}
\label{Eq:SoSta_eisen_eigenmodes}
\psi_k^{\hbox{\scriptsize s,c}}(\vec x\,) \; = \;
\frac 1{\sqrt{\pi a b}} \, x_3 \,
\left( \begin{tabular}[c]{c} $\sin(k \ln x_3)$ \\
$\cos(k \ln x_3)$ \end{tabular} \right)
\hspace{10pt} .
\end{equation}
These plane wave solutions are normalized similar to
(\ref{Eq:SoSta_cusp_eigenmodes}) as
$$
\int_{\cal F} d\mu \,
\psi_k^{\hbox{\scriptsize s,c}}(\vec x\,) \,
\psi_{k'}^{\hbox{\scriptsize s,c}}(\vec x\,) \; = \; \delta(k-k')
\hspace{10pt} .
$$
The metric perturbation can be expanded analogously to
(\ref{Eq:Metric_expansion_SoSta_cusp}), but without the $m$ and
$n$ summation
\begin{equation}
\label{Eq:Metric_expansion_SoSta_sin_cos}
\Phi^{\hbox{\scriptsize s,c}}(\eta,\vec x\,) \; = \;
\int_0^\infty dk \; r(k) \; \Phi_k(\eta) \;
\psi^{\hbox{\scriptsize s,c}}_k(\vec x\,)
\hspace{10pt} ,
\end{equation}
where $r(k)$ are real Gaussian expansion coefficients.

These solutions were not taken into account in the computation of
the density perturbations in
\cite{Sokolov_Starobinskii_1976,Levin_Barrow_Bunn_Silk_1997}.
In \cite{Sokolov_Starobinskii_1976} it was argued
that they should be omitted since they are not bounded as
$x_3\to \infty$.
Constructing small perturbations from (\ref{Eq:SoSta_eisen_eigenmodes})
would then require a fine-tuned superposition of these solutions,
which would contradict the statistical hypothesis.

Here, a remark is in order concerning the numerical evaluation of the
$k$-integrals (\ref{Eq:Metric_expansion_SoSta_cusp}) and
(\ref{Eq:Metric_expansion_SoSta_sin_cos}) in the case that
the integrands contain a Gaussian random function $r(k,m,n)$ or $r(k)$.
In the case of a discrete $k$-spectrum there is no problem.
The integral is then replaced by a sum,
as in equation (\ref{Eq:Metric_expansion}), and each mode $k_n$ occurring
in the spectrum comes with an independent Gaussian random variable.
But in the case of a continuous $k$-spectrum there arises the
question of how small the step size has to be in order to approximate
the integral numerically.
Since the integrand is erratic, one cannot rely on the usual arguments
concerning an approximation as, say, a Riemannian sum.
Even infinitesimally close values of $k$ possess uncorrelated
coefficients due to the Dirac-delta $\delta(k-k')$ in the
correlation function $\left< r(k) r(k') \right>$.
E.\,g.\ consider a unit interval $\Delta k = 1$ and approximate the
integral on $\Delta k$ by $N$ terms.
It is clear that in the limit $N\to\infty$ the approximation to the integral
approaches zero due to the increasing number of cancellations
among the random terms.
This implies that the normalization factor $\alpha$
which is used to normalize the fluctuation to the amplitude of
COBE or WMAP depends on the discretization
used in the approximation of the integrals.
The value of $N$ has at least to be so large that there are many
evaluation points on an interval
where the integrand without the Gaussian random variable is nearly constant.
Consider the mode $\psi_k^{\hbox{\scriptsize s}}(\vec x\,)$,
eq.\,(\ref{Eq:SoSta_eisen_eigenmodes}), having the simple $x_3$-dependence
$x_3 \sin(k \ln x_3)$.
This implies that in order to simulate a Gaussian random field
there is for a fixed $N$ a maximal value of $x_3$
such that there are enough evaluation points for each sine-oscillation.
Thus one has to ensure that there are no parts of the SLS so high
in the horn, i.\,e.\ have corresponding high values of $x_3$,
that the CMB simulation is not valid for the chosen value of $N$.
We use in the following simulations a Gaussian quadrature with
16 evaluation points on the unit interval $\Delta k = 1$.
This is sufficient for the chosen cosmological parameters.
Finally, let us remark that, if there are several contributing integrals,
their normalization factors can be different if their integrands
without the random variable oscillate on different spatial scales.

As a first application let us consider only the decaying modes
(\ref{Eq:SoSta_cusp_eigenmodes}) as it is done in
\cite{Sokolov_Starobinskii_1976,Levin_Barrow_Bunn_Silk_1997}.
In \cite{Levin_Barrow_Bunn_Silk_1997} the temperature fluctuations
$\delta T/T$ are computed for the Sokolov-Starobinskii model
and it is claimed that the periodic horn topology produces
a flat spot in the sky map of the cosmic microwave radiation.
The flat spot, i.\,e.\ a negligible perturbation in the horn,
is claimed to be the result of the asymptotic behaviour of
the eigenmodes (\ref{Eq:SoSta_cusp_eigenmodes})
which are exponentially declining for $z = Q x_3 \gtrsim k$,
because of $K_{\is k}(z) \simeq \sqrt{\frac \pi{2z}} e^{-z}$ in this range.
Modes with $k \lesssim 2\pi x_3$ give a negligible contribution at $x_3$
where the smallest possible value for $Q$,
i.\,e.\ $Q=2\pi$ in the case $a=b=1$, is considered,
since these are the modes which reach as far as possible into the horn.
In \cite{Levin_Barrow_Bunn_Silk_1997} the cut-off in $k$ is so low
that the modes cannot produce a perturbation in the horn
at the position which corresponds to the distance of the
surface of last scattering.
As a consequence, the sky maps computed in \cite{Levin_Barrow_Bunn_Silk_1997}
do not show temperature fluctuations in the horn.
Increasing, however, the cut-off from the low value $k_c=10$ used in
\cite{Levin_Barrow_Bunn_Silk_1997} to the value $k_c=140$
shows that there are indeed fluctuations in the horn
as it will be demonstrated below.

\begin{figure}[htb]
\begin{center}
\hspace*{-10pt}\begin{minipage}{9cm}
\includegraphics[width=9.0cm]{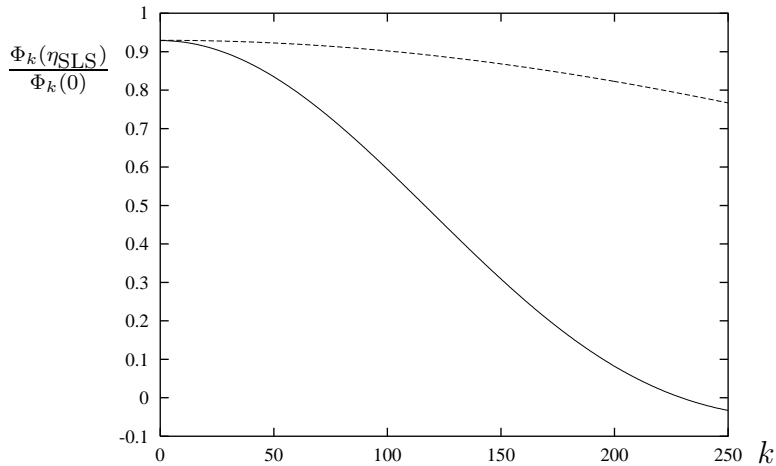}
\put(-1,4){$k$}
\put(-285,150){$\frac{\Phi_k(\eta_{\hbox{\scriptsize{SLS}}})}{\Phi_k(0)}$}
\end{minipage}
\vspace*{-10pt}
\end{center}
\caption{\label{Fig:f_k_eta_sls}
The $k$-dependence of $\Phi_k(\eta_{\hbox{\scriptsize{SLS}}})/\Phi_k(0)$
in the single-fluid approximation is shown
for a model including radiation with $\Omega_{\hbox{\scriptsize mat}} = 0.3$
and $\Omega_\Lambda = 0$ (full curve) and a nearly flat one with
$\Omega_{\hbox{\scriptsize mat}} = 0.3$ and
$\Omega_\Lambda = 0.65$ (dashed curve).}
\end{figure}

There is nevertheless a subtle effect which can still suppress
the amplitude of the perturbations in the horn even in the case of
a high cut-off $k_c$.
As the above discussion has emphasized, it is important
to include modes with sufficiently high values of $k$.
If, however, the time evolution of $\Phi_k(\eta)$ leads
to a too strong suppression of these modes at the surface of last scattering,
i.\,e.\ if $\Phi_k(\eta_{\hbox{\scriptsize{SLS}}})$ declines
as a function of $k$ too fast at the values of $k$
which generate anisotropy in the horn,
then these modes cannot contribute as much as they would for
a nearly $k$-independent $\Phi_k(\eta_{\hbox{\scriptsize{SLS}}})$.
A pure matter model is considered in \cite{Levin_Barrow_Bunn_Silk_1997}
for which the time evolution of $\Phi_k(\eta)/\Phi_k(0)$
is independent of $k$
\begin{equation}
\label{Eq:f_k_Matter}
\Phi_k^{\hbox{\scriptsize mat}}(\eta) \; = \;
\Phi_k^{\hbox{\scriptsize mat}}(0) \;
\frac{5(\sinh^2\eta-3\eta\sinh\eta+4\cosh\eta-4)}{(\cosh\eta-1)^3}
\hspace{10pt} .
\end{equation}
Therefore this effect does not occur in the calculation
in \cite{Levin_Barrow_Bunn_Silk_1997}.
In general, however, especially if one takes also radiation into account,
equation (\ref{Eq:Phi_first-order}) leads to
a $k$-dependent time evolution of $\Phi_k(\eta)/\Phi_k(0)$ as can be seen
in figure \ref{Fig:f_k_eta_sls}
where $\Phi_k(\eta_{\hbox{\scriptsize{SLS}}})/\Phi_k(0)$ is shown for
two models containing also radiation,
one with $\Omega_{\hbox{\scriptsize mat}} = 0.3$ and
$\Omega_\Lambda = 0$, and a nearly flat one with
$\Omega_{\hbox{\scriptsize mat}} = 0.3$ and
$\Omega_\Lambda = 0.65$.
The model with $\Omega_{\hbox{\scriptsize tot}} = 0.3$ has a much
stronger $k$-dependence as the nearly flat model.
Notice that for a pure matter model one has
$\Phi_k(\eta_{\hbox{\scriptsize{SLS}}})/\Phi_k(0) = \hbox{const}$.
The value of $\Phi_k(\eta)$ at $\eta_{\hbox{\scriptsize{SLS}}}$ plays an
important role because it describes the metric perturbation at recombination
and via eq.\ (\ref{Eq:Sachs_Wolfe}) the CMB anisotropy.

\begin{figure}[htb]
\begin{center}
\hspace*{-150pt}\begin{minipage}{7cm}
\includegraphics[width=7.0cm,angle=270]{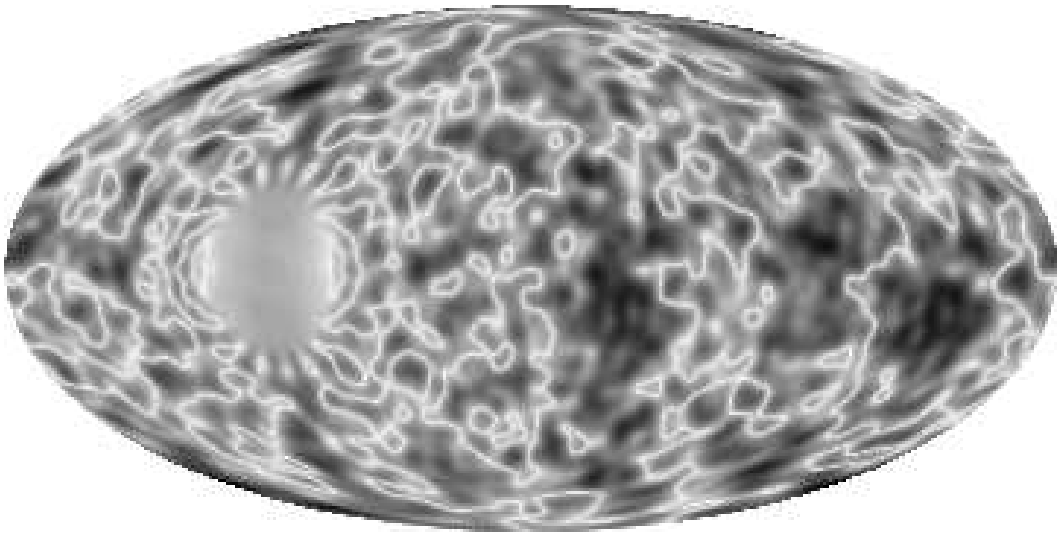}
\end{minipage}
\vspace*{-10pt}
\end{center}
\caption{\label{Fig:Horned_k_010_l_70_kbes_HZ_ran}
The CMB anisotropy $\delta T$ for the Sokolov-Starobinskii model for
$\Omega_{\hbox{\scriptsize mat}} = 0.3$ and
$\Omega_\Lambda = 0$ with a cut-off $k_c=10$
using a Gaussian superposition of the modes (\ref{Eq:SoSta_cusp_eigenmodes}).
The white ``curves'' belong to $\delta T=0$ whereas increasing
darkness corresponds to increasing deviations from $\delta T=0$.
}
\end{figure}

\begin{figure}[htb]
\begin{center}
\hspace*{-150pt}\begin{minipage}{7cm}
\includegraphics[width=7.0cm,angle=270]{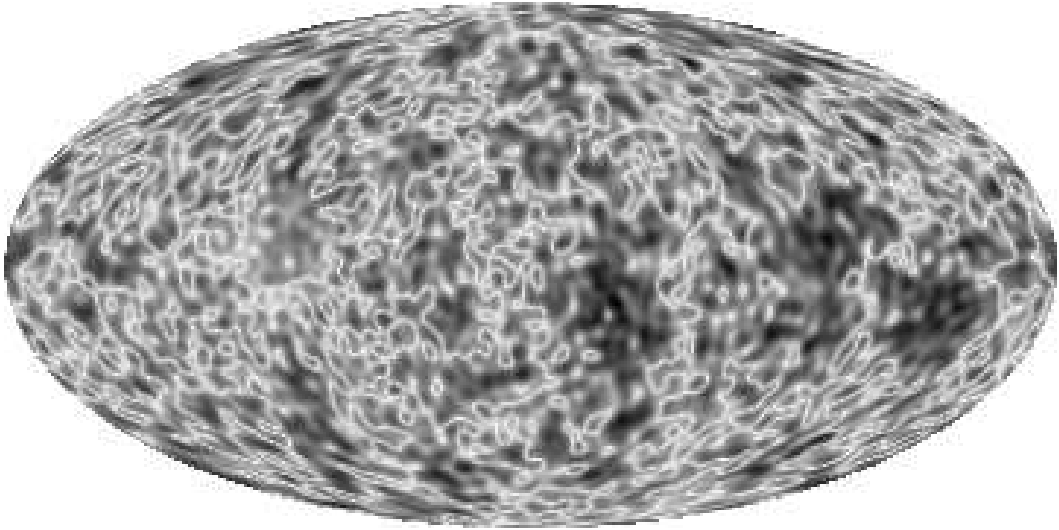}
\end{minipage}
\vspace*{-10pt}
\end{center}
\caption{\label{Fig:Horned_k_140_l_70_kbes_HZ_ran}
The CMB anisotropy $\delta T$ for the Sokolov-Starobinskii model for
$\Omega_{\hbox{\scriptsize mat}} = 0.3$ and
$\Omega_\Lambda = 0$ with a cut-off $k_c=140$
using a Gaussian superposition of the modes (\ref{Eq:SoSta_cusp_eigenmodes}).
}
\end{figure}

\begin{figure}[htb]
\begin{center}
\hspace*{-10pt}\begin{minipage}{9cm}
\includegraphics[width=9.0cm]{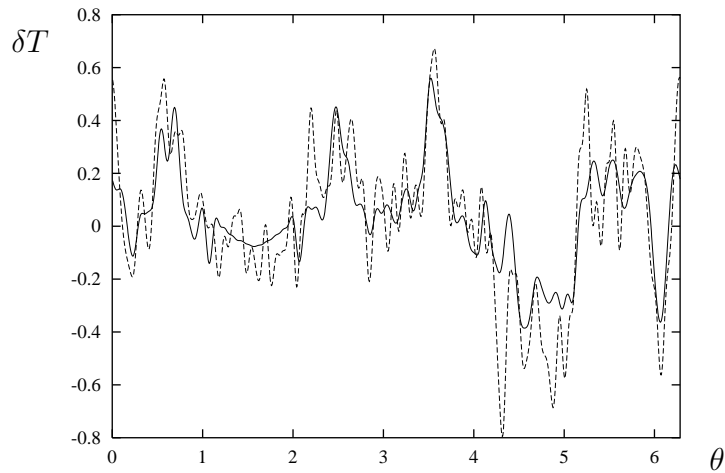}
\put(-1,2){$\theta$}
\put(-265,160){$\delta T$}
\end{minipage}
\vspace*{-10pt}
\end{center}
\caption{\label{Fig:SoSta_CMB_Meridian_cusp_k_10_140}
The CMB anisotropy $\delta T$ along the equator of
figures \ref{Fig:Horned_k_010_l_70_kbes_HZ_ran}
and \ref{Fig:Horned_k_140_l_70_kbes_HZ_ran}.
The suppression in the direction of the horn at $\theta=\frac \pi 2$
is clearly visible in the case of the $k_c=10$ cut-off (full curve).
In the case of the $k_c=140$ cut-off (dashed curve)
there are now fluctuations towards the horn.
}
\end{figure}

\begin{figure}[htb]
\begin{center}
\hspace*{-10pt}\begin{minipage}{9cm}
\includegraphics[width=9.0cm]{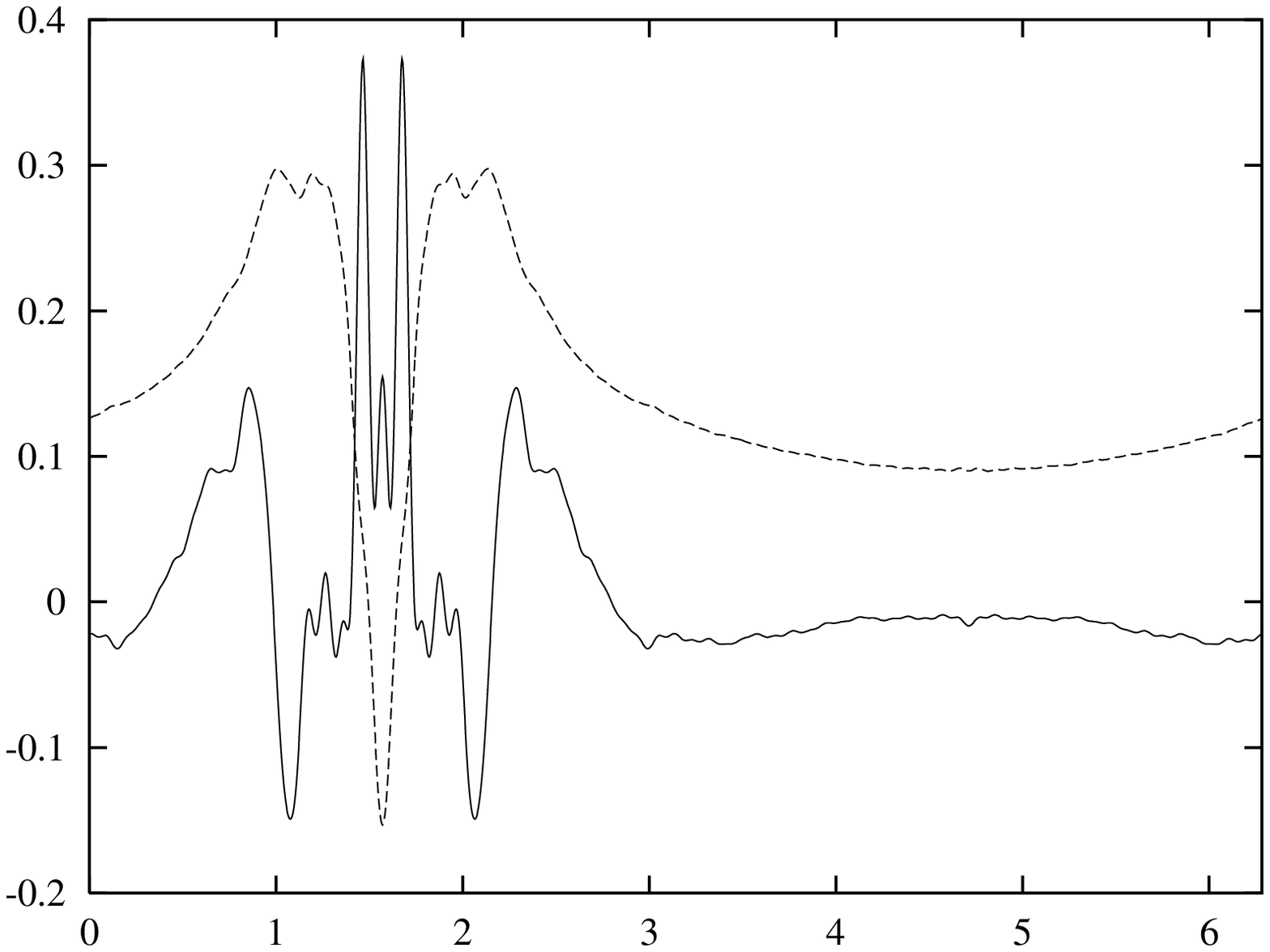}
\put(-1,2){$\theta$}
\put(-265,160){$\delta T$}
\end{minipage}
\vspace*{-10pt}
\end{center}
\caption{\label{Fig:SoSta_CMB_Meridian_sin_cos}
The CMB anisotropy $\delta T$ along the equator of
the sine (full curve) and cosine (dashed curve) plane wave solutions
(\ref{Eq:SoSta_eisen_eigenmodes}) using a Gaussian superposition.
The contribution from the cosine plane waves is multiplied by a factor
of $-\frac 16$ in order to get fluctuations
comparable to the sine plane wave contribution.
}
\end{figure}

Let us now discuss the cosmic microwave anisotropy of the
Sokolov-Starobinskii model,
which is here computed using the tight-coupling approximation.
In contrast to \cite{Levin_Barrow_Bunn_Silk_1997},
we are not using a pure matter model and include the
naive and integrated Sachs-Wolfe effect as well as the
Doppler effect.
The observer is located at $(x_1,x_2,x_3)=(0,0,1)$.
For the coefficients $r(k,m,n)$ Gaussian random variables
are chosen.
To illustrate the dependence on the numerically chosen cut-off in $k$,
consider at first a model with
$\Omega_{\hbox{\scriptsize mat}} = \Omega_{\hbox{\scriptsize tot}} = 0.3$
as in \cite{Levin_Barrow_Bunn_Silk_1997}.
In figure \ref{Fig:Horned_k_010_l_70_kbes_HZ_ran}
the CMB anisotropy is computed using the very low cut-off at $k_c=10$.
The coordinate system is chosen such that the horn lies at the
equator at one fourth from the left.
One observes a suppression of the anisotropy in the horn as discussed above.
In figure \ref{Fig:Horned_k_140_l_70_kbes_HZ_ran} the cut-off is
increased to $k_c=140$ and no suppression is observed.
The figure \ref{Fig:SoSta_CMB_Meridian_cusp_k_10_140} represents
the anisotropy $\delta T$ along the equator in the
coordinate system of figures \ref{Fig:Horned_k_010_l_70_kbes_HZ_ran} and
\ref{Fig:Horned_k_140_l_70_kbes_HZ_ran}
(the horizontal line bisecting the sky map).
The horn lies at $\theta=\frac \pi 2$.
The full curve shows the anisotropy for a cut-off at $k_c=10$
revealing clearly a suppression of the anisotropy around $\theta=\frac \pi 2$.
The dashed curve presenting the $k_c=140$ cut-off
displays anisotropies also in the horn.

\begin{figure}[htb]
\begin{center}
\hspace*{-150pt}\begin{minipage}{7cm}
\includegraphics[width=7.0cm,angle=270]{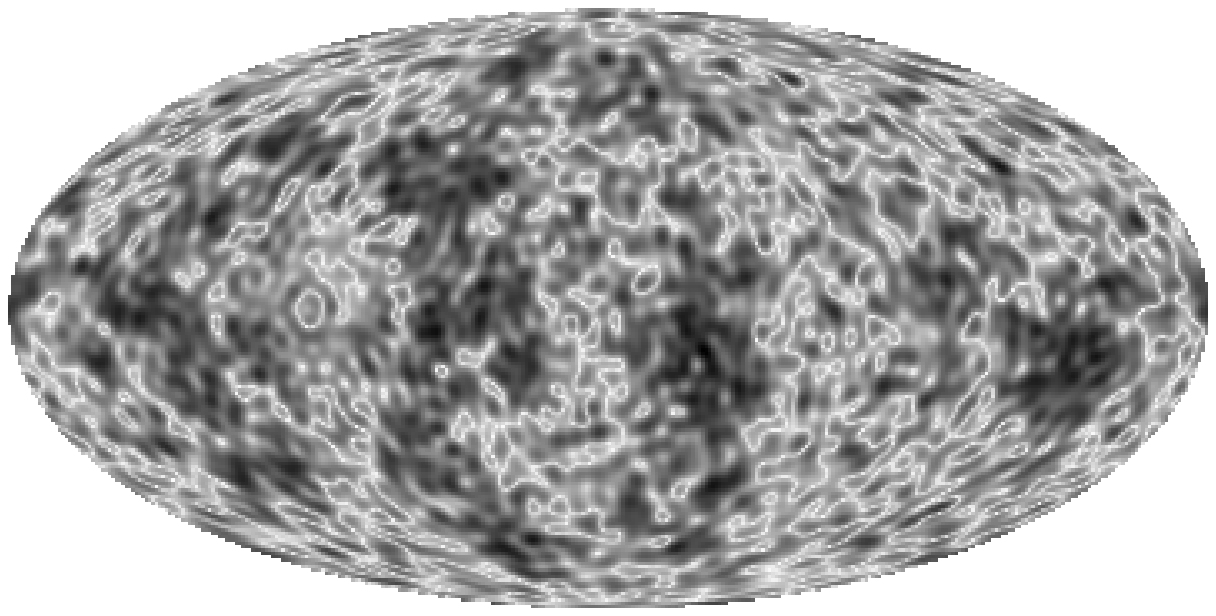}
\end{minipage}
\vspace*{-10pt}
\end{center}
\caption{\label{Fig:Horned_k_140_l_70_merge3_ran}
The CMB anisotropy $\delta T$ for the Sokolov-Starobinskii model for
$\Omega_{\hbox{\scriptsize mat}} = 0.3$ and
$\Omega_\Lambda = 0$ with a cut-off $k_c=140$.
A Gaussian superposition of all three types of modes is shown,
i.\,e.\ the modes (\ref{Eq:SoSta_cusp_eigenmodes}) and the sine and cosine
plane waves (\ref{Eq:SoSta_eisen_eigenmodes}).
The contribution from the cosine plane waves is weighted by a factor
of $-\frac 16$.
}
\end{figure}

\begin{figure}[htb]
\begin{center}
\hspace*{-10pt}\begin{minipage}{9cm}
\includegraphics[width=9.0cm]{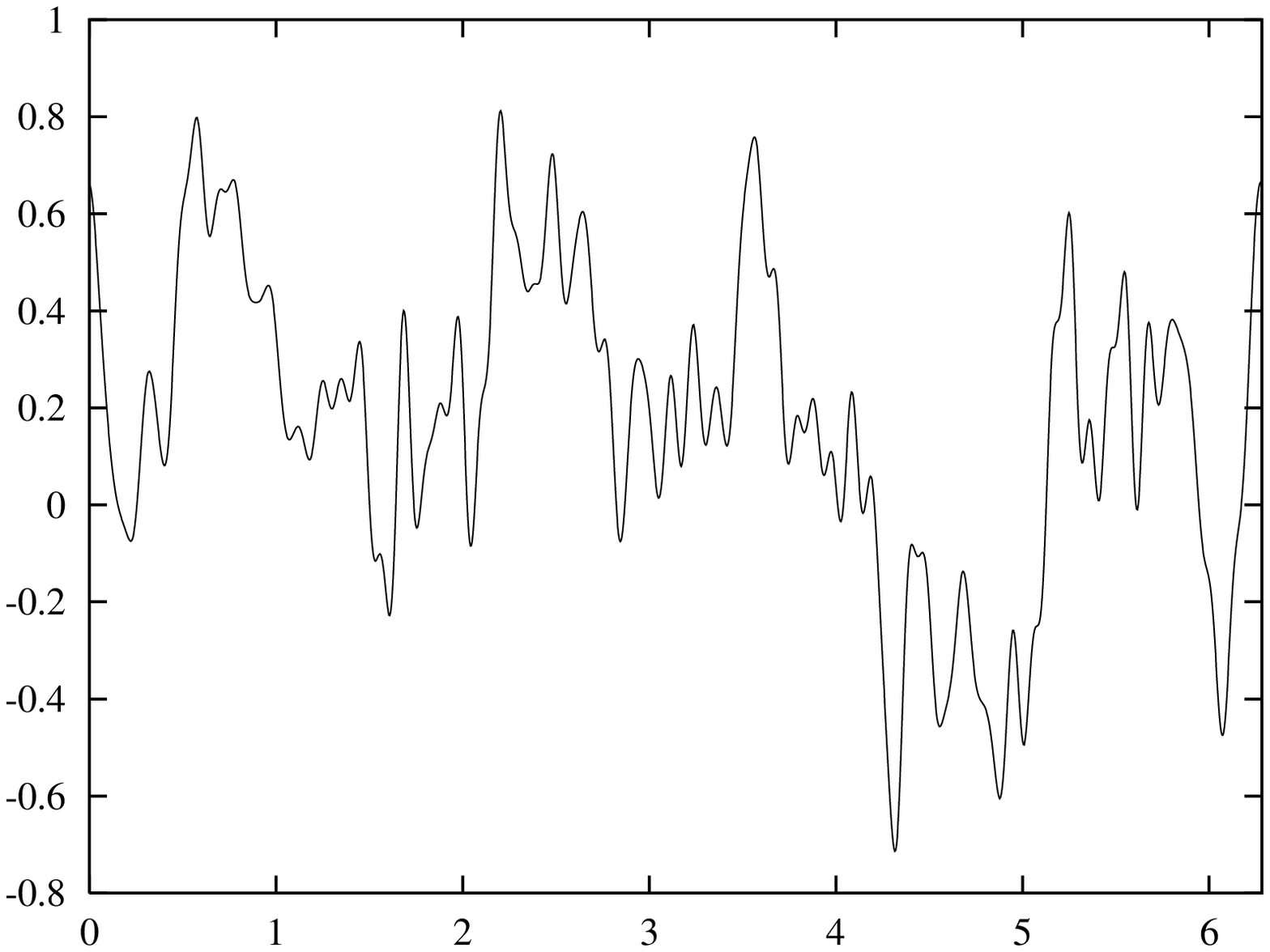}
\put(-1,4){$\theta$}
\put(-260,155){$\delta T$}
\end{minipage}
\vspace*{-10pt}
\end{center}
\caption{\label{Fig:SoSta_Merge3_equator}
The CMB anisotropy is shown at the equator of the sky map from figure
\protect\ref{Fig:Horned_k_140_l_70_merge3_ran}.
}
\end{figure}

Up to now we have computed the CMB anisotropy solely with
the eigenmodes (\ref{Eq:SoSta_cusp_eigenmodes}) and
have ignored the plane wave solutions (\ref{Eq:SoSta_eisen_eigenmodes}).
These solutions contribute significantly towards the horn.
As can be observed from figure \ref{Fig:SoSta_CMB_Meridian_cusp_k_10_140}
there are fluctuations for a sufficiently high chosen cut-off in $k$,
but these fluctuations are smaller than farther away from the horn.
This is due to the current restriction to the 
eigenmodes (\ref{Eq:SoSta_cusp_eigenmodes}).
The plane wave solutions have significant contributions especially
in the direction of the horn
as shown in figure \ref{Fig:SoSta_CMB_Meridian_sin_cos}.
We now compute the CMB anisotropy for the model with
$\Omega_{\hbox{\scriptsize tot}} = 0.3$ and a cut-off $k_c=140$
using all three types of modes.
We use for the coefficients $r(k,m,n)$ Gaussian random variables.
The result is presented in figure \ref{Fig:Horned_k_140_l_70_merge3_ran}
showing the CMB sky map and in figure \ref{Fig:SoSta_Merge3_equator}
showing the anisotropy along the equator.
As can be seen in figures \ref{Fig:Horned_k_140_l_70_merge3_ran} and
\ref{Fig:SoSta_Merge3_equator},
taking the sine and cosine plane waves into account increases
the amplitude of the fluctuations towards the horn.
Thus the fluctuations do not betray the horned topology.
Since we use Gaussian random variables as coefficients,
there is also no fine tuning in contrast to what is stated
in \cite{Sokolov_Starobinskii_1976}.

\begin{figure}[htb]
\begin{center}
\hspace*{-10pt}\begin{minipage}{9cm}
\includegraphics[width=9.0cm]{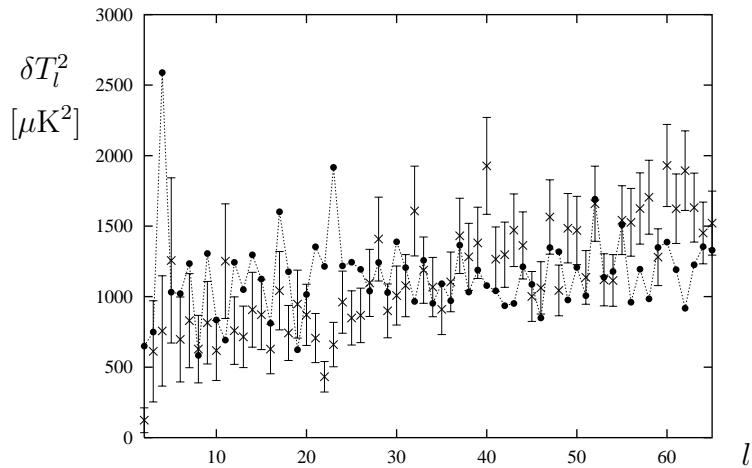}
\put(-1,4){$l$}
\put(-274,150){$\delta T_l^2$}
\put(-278,130){$[\mu \hbox{K}^2]$}
\end{minipage}
\vspace*{-10pt}
\end{center}
\caption{\label{Fig:SoSta_Merge3_Cl}
The angular power spectrum $\delta T_l^2 = l(l+1) C_l/2\pi$ is
shown for the sky map from figure
\protect\ref{Fig:Horned_k_140_l_70_merge3_ran} as dots.
The cruxes with error bars are the first year WMAP data
\cite{Hinshaw_et_al_2003}.
}
\end{figure}

After having shown that the horn can be masked by the CMB fluctuations,
let us now turn to the statistical properties of the
CMB anisotropy.
Expanding the temperature fluctuations with respect
to spherical harmonics $Y_l^m$ yields
the expansion coefficients $a_{lm}$
which in turn lead to the multipoles $C_l$, see eq.\,(\ref{Eq:C_l}).
The angular power spectrum $\delta T_l^2 = l(l+1) C_l/2\pi$
is shown in figure \ref{Fig:SoSta_Merge3_Cl}
for the model where the superposition of all modes is used,
i.\,e.\ for the anisotropy shown
in figure \ref{Fig:Horned_k_140_l_70_merge3_ran}.
Only large angular scales, i.\,e.\ multipoles with $l < 65$, are shown
since higher values of $l$ would require modes above our
numerical cut-off at $k_c=140$.
The values of $\delta T_l^2$ display a flat spectrum up to $l\simeq 40$
as it is also revealed by the WMAP data \cite{Hinshaw_et_al_2003}.
(The error bars of the WMAP data shown in figure \ref{Fig:SoSta_Merge3_Cl}
do not include the cosmic variance since we compare one-sky realizations.)
The large fluctuations of $\delta T_l^2$ in our model
arise from the fact that we compute these values for a fixed observer
and not, like CMBFast or CAMB, as a statistical average.
The WMAP data display a very low quadrupole moment
which is not reproduced by this simulation.
This is due to the infinite volume of the Sokolov-Starobinskii
fundamental cell.
This contrasts to the Picard cell which is discussed in the next section.

\begin{figure}[htb]
\begin{center}
\hspace*{-10pt}\begin{minipage}{9cm}
\includegraphics[width=9.0cm]{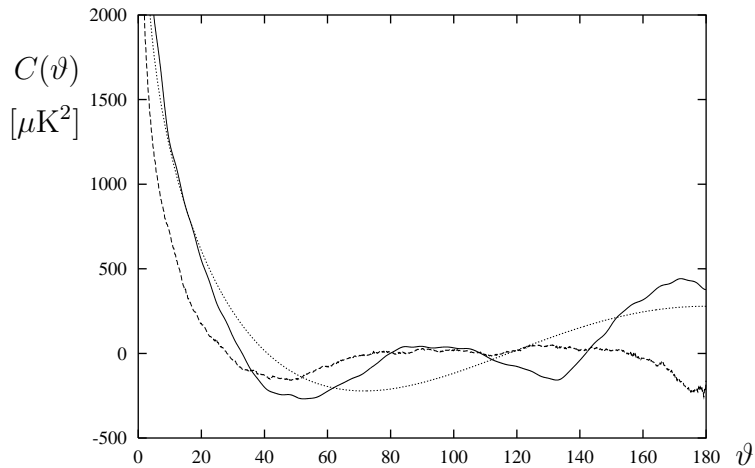}
\put(-1,4){$\vartheta$}
\put(-274,150){$C(\vartheta)$}
\put(-276,130){$[\mu \hbox{K}^2]$}
\end{minipage}
\vspace*{-10pt}
\end{center}
\caption{\label{Fig:SoSta_Merge3_C_theta}
The temperature correlation function $C(\vartheta)$
is shown as a full curve for the sky map from figure
\protect\ref{Fig:Horned_k_140_l_70_merge3_ran}.
The corresponding WMAP curve is displayed as a dashed curve.
The dotted curve represents the best $\Lambda$CDM model
described in \cite{Bennett_et_al_2003}.
}
\end{figure}

Let us now turn to the correlation function $C(\vartheta)$
which emphasizes the large angular scales and thus the low $l$-range.
In figure \ref{Fig:SoSta_Merge3_C_theta} $C(\vartheta)$ is shown
(full curve) in comparison with the WMAP data \cite{Hinshaw_et_al_2003}
(dashed curve).
A large deviation from the WMAP data is revealed near $\vartheta=0^\circ$
since our computation takes only the values of $C_l$ with $l\leq 65$
into account.
The model does not describe the correlation hole for
$\vartheta \gtrsim 160^\circ$
as observed by WMAP which is due to the very small observed
quadrupole moment $C_2$.

\begin{figure}[htb]
\begin{center}
\hspace*{33pt}\begin{minipage}{14cm}
\includegraphics[width=14.0cm]{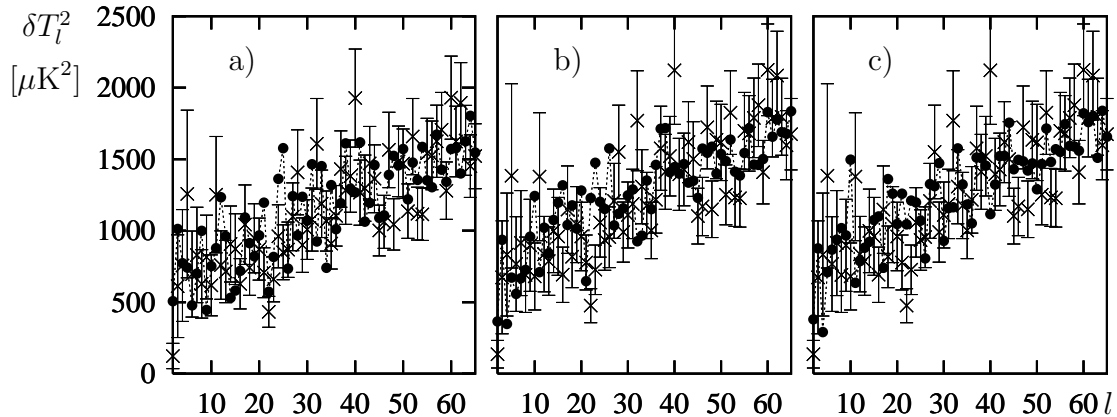}
\put(-1,7){$l$}
\put(-410,150){$\delta T_l^2$}
\put(-415,130){$[\mu \hbox{K}^2]$}
\put(-332,137){a)}
\put(-209,137){b)}
\put(-90,137){c)}
\end{minipage}
\vspace*{-10pt}
\end{center}
\caption{\label{Fig:SoSta_Cl}
The angular power spectrum $\delta T_l^2 = l(l+1) C_l/2\pi$ is shown
(full dots) for three nearly flat Sokolov-Starobinskii models having
a) $\Omega_{\hbox{\scriptsize mat}} = 0.3$,  $\Omega_\Lambda = 0.6$,
b) $\Omega_{\hbox{\scriptsize mat}} = 0.3$,  $\Omega_\Lambda = 0.65$ and
c) $\Omega_{\hbox{\scriptsize mat}} = 0.35$,  $\Omega_\Lambda = 0.6$.
The cruxes with error bars are the first year WMAP data
\cite{Hinshaw_et_al_2003}.
}
\end{figure}

Up to now we have only discussed a model with
$\Omega_{\hbox{\scriptsize tot}} = 0.3$.
However, current cosmological observations now point to a nearly flat
universe.
Therefore we show in figure \ref{Fig:SoSta_Cl} angular power spectra
$\delta T_l^2$ for models with $\Omega_{\hbox{\scriptsize tot}} = 0.9$
and $\Omega_{\hbox{\scriptsize tot}} = 0.95$.
We use here only the eigenmodes (\ref{Eq:SoSta_cusp_eigenmodes})
up to $k_c=140$ and ignore the plane waves.
In order to obtain the correct values for higher values of $l$
the contribution of modes above $k=140$ is taken into account by assuming
that their spherical expansion with respect to the observer point
yields Gaussian expansion coefficients $a_{lm}$,
i.\,e.\ that they are statistically isotropic.
Under this assumption, the $k$-integral in (\ref{Eq:Sachs_Wolfe_tight_coupling})
is for $k>140$ evaluated analogously to the CMBFast method.
In figure \ref{Fig:SoSta_Cl}a) we choose
$\Omega_{\hbox{\scriptsize mat}} = 0.3$,  $\Omega_\Lambda = 0.6$,
in figure \ref{Fig:SoSta_Cl}b)
$\Omega_{\hbox{\scriptsize mat}} = 0.3$,  $\Omega_\Lambda = 0.65$ and
in figure \ref{Fig:SoSta_Cl}c)
$\Omega_{\hbox{\scriptsize mat}} = 0.35$,  $\Omega_\Lambda = 0.6$.
In all three cases an approximately flat spectrum is observed having
up to values of $l=65$ fluctuations of the same order as the WMAP data.
The topological structure of the Sokolov-Starobinskii model
produces no signature in the angular power spectrum.
Thus if our Universe would possess a small negative curvature with
the Sokolov-Starobinskii topology,
the CMB anisotropy would not reveal this ``horned'' topology.

\section{The Picard model}

\begin{figure}
\includegraphics[width=5cm,height=8cm,angle=-90]{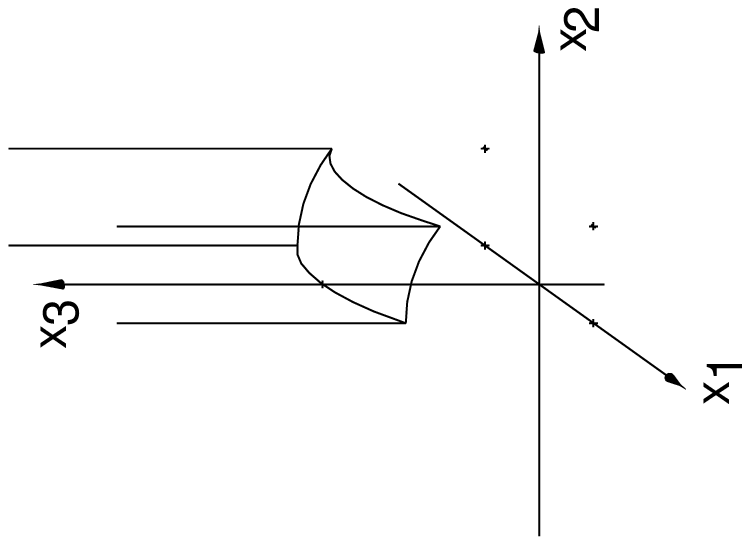}
\hspace*{-60pt}\includegraphics[width=5cm,height=8cm,angle=-90]{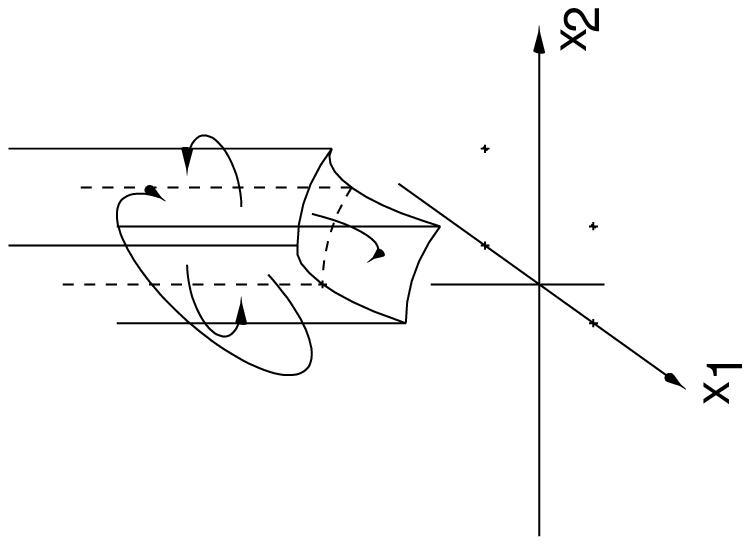}
\put(-325,-30){a)}
\put(-170,-30){b)}
\caption{\label{Fig:Fundpoly3d}
The fundamental domain of the Picard group and the identifications
of the surfaces.}
\end{figure}

The crucial difference between the Sokolov-Starobinskii model and
the Picard model \cite{Picard_1884} is
that the volume of the Picard model is finite.
This is achieved by cutting off the lower $(x_3\to 0)$ part of the
Sokolov-Starobinskii model which leads to the infinite volume.
In the Picard model one chooses $a=1$ and $b=1$ for the
periodical identification in the $x_1$- and $x_2$-directions,
respectively.
In addition to these two identifications, in the Picard model
all points are identified with those
which are inverted at the unit sphere having radius 1
and its center at $\vec x = (0,0,0)$.
To describe this transformation let us represent the points
$\vec x \in \HS$ by Hamilton quaternions,
$q = x_1 + \i x_2 + \j x_3$,
with the multiplication defined by $\i^2=-1, \j^2=-1, \i\j+\j\i=0$
plus the property that $\i$ and $\j$ commute with every real number.
The inverse of a quaternion
$q=q_1+q_2\i+q_3\j+q_4\i\j\neq 0$ is then given by
$q^{-1}=|q|^{-2}(q_1-q_2\i-q_3\j-q_4\i\j)$,
where $|q|^2=q_1^2+q_2^2+q_3^2+q_4^2$.
The Picard group $\Gamma$ is then generated by
two translations and one inversion,
\begin{equation}
\label{Eq:Picard_Generators}
q\mapsto q+1, \quad q\mapsto q+\i, \quad q\mapsto-q^{-1}
\hspace{10pt} .
\end{equation}
The transformations $\gamma$ in $\HS$ are given by
linear fractional transformations
$$
q\mapsto \gamma \, q = (aq+b) (cq+d)^{-1}
\hspace{10pt} ; \hspace{10pt}
a,b,c,d \in \mathbb{C}
\hspace{10pt} ; \hspace{10pt}
ad-bc = 1
\hspace{10pt} .
$$
The group of these transformations is up to a common sign
isomorphic to the group of matrices of PSL($2,\mathbb{C}$).
The inversion $q\mapsto-q^{-1}$ bounds the fundamental domain
by the unit sphere from below.
From the three generating transformations (\ref{Eq:Picard_Generators})
one can compose the transformation $q \mapsto \i q \i$
which corresponds to a rotation by 180$^\circ$ around
the $x_3$-axis.
This rotation together with the inversion $q\mapsto-q^{-1}$
are the new identifications in comparison with
the Sokolov-Starobinskii model.
The fundamental domain of standard shape is then given by
\cite{Picard_1884,Elstrodt_Grunewald_Mennicke_1998}
\begin{equation}
\label{Picard_Fundamental_Cell}
{\cal F}=\left\{ q=x_1 + \i x_2+\j x_3 \Big|\;
-\frac 12 <x_1<\frac 12, \; 0<x_2<\frac 12, \; |q|>1\right\}
\hspace{4pt} ,
\end{equation}
see also figure \ref{Fig:Fundpoly3d}.
The fundamental domain ${\cal F}$ is a hyperbolic pyramid with one vertex
at $\infty$ and the other four vertices  in the points
$P_1 = -\frac 12 + \frac{\sqrt 3}2 \j$, 
$P_2 =  \frac 12 + \frac{\sqrt 3}2 \j$, 
$P_3 =  \frac 12 + \frac 12 \i +  \frac{\sqrt 2}2 \j$, and
$P_4 = -\frac 12 + \frac 12 \i +  \frac{\sqrt 2}2 \j$.
The volume of this non-compact orbifold 
$\Gamma\backslash\HS$ is finite \cite{Humbert_1919},
\begin{equation}
\hbox{vol}(\Gamma\backslash \HS) \; = \; \frac{\zeta_K(2)}{4\pi^2}
\; = \; 0.30532186\dots
\end{equation}
where
$$
\zeta_K(s) \; = \; \frac 14\sum_{\nu\in\mathbb{Z}[\is]-\{0\}}
(\nu\bar{\nu})^{-s}, \quad \Re s>1
\hspace{10pt} ,
$$
is the Dedekind zeta function,
and $\mathbb{Z}[\i] = \mathbb{Z} + \i \mathbb{Z}$ are the Gaussian integers.

Due to the additional inversion, the non-trivial eigenfunctions
of the Laplace-Beltrami operator are more complicated than in the
Sokolov-Starobinskii case and are not known analytically.
The eigenfunctions are the so-called Maa\3 waveforms \cite{Maass_1949b}
which are automorphic, i.\,e.\ satisfy
$\psi(\gamma q)=\psi(q) \; \forall \; \gamma \in \Gamma, \; q\in\HS$,
and therefore periodic in $x_1$ and $x_2$.
It follows then that they can be expanded into a Fourier series
with respect to $x_1$ and $x_2$
\begin{equation}
\label{FourierExpansion_Picard}
\psi_k(\vec x\,) \; = \; u(x_3) \, + \,
\sum_{(m+\is n)\in\mathbb{Z}[\is]-\{0\}}
a_{kmn} \, x_3 \, K_{\is k}(Qx_3) \; e^{2\pi\is(m x_1+n x_2)}
\hspace{10pt} ,
\end{equation}
where $Q = 2\pi\sqrt{m^2+n^2}$ and
\begin{equation}
u(x_3) \, = \,
\cases{
b_0 x_3^{1+\is k}+b_1 x_3^{1-\is k} & \hbox{if $k\not=0$}\\
b_2 x_3 + b_3 x_3\ln x_3 & \hbox{if $k=0$}\\}
\hspace{15pt} .
\end{equation}
(Here we write $\psi(\vec x\,)=\psi(x_1,x_2,x_3)$ instead of $\psi(q)$.)
$K_{\is k}(z)$ denotes again the $K$-Bessel function
whose order is connected with the eigenvalue $E$ by $E=k^2+1$.
If a Maa\3 waveform vanishes in the cusp,
\begin{equation}
\lim_{x_3\to \infty} \, \psi_k^{\hbox{\scriptsize cusp}}(\vec x\,) \; = \; 0
\hspace{10pt} ,
\end{equation}
it is called a Maa\3 cusp form.
Maa\3 cusp forms have
$b^{\hbox{\scriptsize cusp}}_0=b^{\hbox{\scriptsize cusp}}_1=
b^{\hbox{\scriptsize cusp}}_2=b^{\hbox{\scriptsize cusp}}_3=0$
and are square integrable over the fundamental domain,
$\langle \psi_k^{\hbox{\scriptsize cusp}},
\psi_k^{\hbox{\scriptsize cusp}}\rangle<\infty$, where
\begin{equation}
\langle \psi,\psi'\rangle=\int_{\cal F} d\mu \, \psi \psi'
\hspace{10pt} \hbox{ with } \hspace{10pt}
d\mu \; = \; \frac{dx_1 dx_2 dx_3}{x_3^3}
\hspace{10pt} .
\end{equation}
The coefficients $a^{\hbox{\scriptsize cusp}}_{kmn}$ in
(\ref{FourierExpansion_Picard}) are not free expansion coefficients
as in the Sokolov-Starobinskii model,
but are uniquely determined for each eigenvalue $k$.
This is due to the additional symmetry operation of the Picard cell.

According to the Roelcke-Selberg spectral resolution of the Laplacian
\cite{Selberg_1956,Roelcke_1966,Roelcke_1967}, its spectrum contains both
a discrete and a continuous part.
The discrete part is spanned by the constant eigenfunction
$\psi_{k_0}^{\hbox{\scriptsize cusp}} = \hbox{vol}({\cal F})^{-1/2}$
and a countable number of Maa\3 cusp forms
$\psi_{k_1}^{\hbox{\scriptsize cusp}}$,
$\psi_{k_2}^{\hbox{\scriptsize cusp}}$,
$\psi_{k_3}^{\hbox{\scriptsize cusp}},\ldots$
which we take to be ordered with increasing eigenvalues,
$0=E_0<E_1\le E_2\le E_3\le\ldots$.
The smallest nontrivial eigenvalue is $E_1=43.8522464\dots$ \cite{Steil_1999}.
Recall that the corresponding modes (\ref{Eq:SoSta_cusp_eigenmodes})
in the Sokolov-Starobinskii model possess a continuous spectrum
due to the missing inversion symmetry.

The continuous part of the spectrum $E\ge1$ is spanned by the
Eisenstein series $\psi_k^{\hbox{\scriptsize Eisen}}(\vec x\,)$
which are known analytically
\cite{Kubota_1973,Elstrodt_Grunewald_Mennicke_1985}.
Their Fourier coefficients are given by
\begin{equation}
b^{\hbox{\scriptsize Eisen}}_0 \, = \,
\frac{1}{\sqrt \pi} \, \frac{\Lambda_K(1+\i k)}{|\Lambda_K(1+\i k)|}
\hspace{10pt} , \hspace{10pt}
b^{\hbox{\scriptsize Eisen}}_1 \, = \,
\frac{1}{\sqrt \pi} \,\frac{\Lambda_K(1-\i k)}{|\Lambda_K(1+\i k)|}
\hspace{10pt} ,
\end{equation}
and $b^{\hbox{\scriptsize Eisen}}_2 = b^{\hbox{\scriptsize Eisen}}_3 = 0$ and
\begin{equation}
a^{\hbox{\scriptsize Eisen}}_{kmn} \, = \,
\frac{2}{\sqrt \pi \, |\Lambda_K(1+\i k)|}
\sum_{\lambda,\mu\in\mathbb{Z}[\is],\lambda\mu=m+\is n}
\left| \frac{\lambda}{\mu} \right|^{\is k},
\end{equation}
where
\begin{equation}
\Lambda_K(s) \, = \, 4\pi^{-s} \, \Gamma(s) \, \zeta_K(s)
\end{equation}
has an analytic continuation into the complex plane except for a
pole at $s=1$.
With these coefficients the Eisenstein series are real and normalized
$\langle \psi_k^{\hbox{\scriptsize Eisen}},
\psi_{k'}^{\hbox{\scriptsize Eisen}}\rangle = \delta(k-k')$.
In the case of the Sokolov-Starobinskii model one has
$a_{kmn} = 0$, and $b_0$ and $b_1$ were chosen to obtain the
real plane wave solutions (\ref{Eq:SoSta_eisen_eigenmodes}).

Normalizing the Maa\3 cusp forms according to
$\langle \psi_k^{\hbox{\scriptsize cusp}},
\psi_k^{\hbox{\scriptsize cusp}}\rangle = 1$, 
we can expand any square integrable function
$\phi\in L^2(\Gamma\backslash\HS)$ in terms of Maa\3 waveforms,
\cite{Elstrodt_Grunewald_Mennicke_1998},
\begin{equation}
\label{eq:completeness-relation}
\phi(\vec x\,) \, = \,
\sum_{n\ge0}\langle \psi_{k_n}^{\hbox{\scriptsize cusp}},\phi\rangle \,
\psi_{k_n}^{\hbox{\scriptsize cusp}}(\vec x\,) \, + \,
\int_0^\infty dk \,
\langle \psi_k^{\hbox{\scriptsize Eisen}},\phi\rangle \,
\psi_k^{\hbox{\scriptsize Eisen}}(\vec x\,)
\hspace{10pt} .
\end{equation}
Since the discrete eigenvalues and their associated Maa\3 cusp forms are
not known analytically, one has to compute them numerically.
\begin{figure}[b]
\begin{center}
\hspace*{-150pt}\begin{minipage}{7cm}
\includegraphics[width=7.0cm,angle=270]{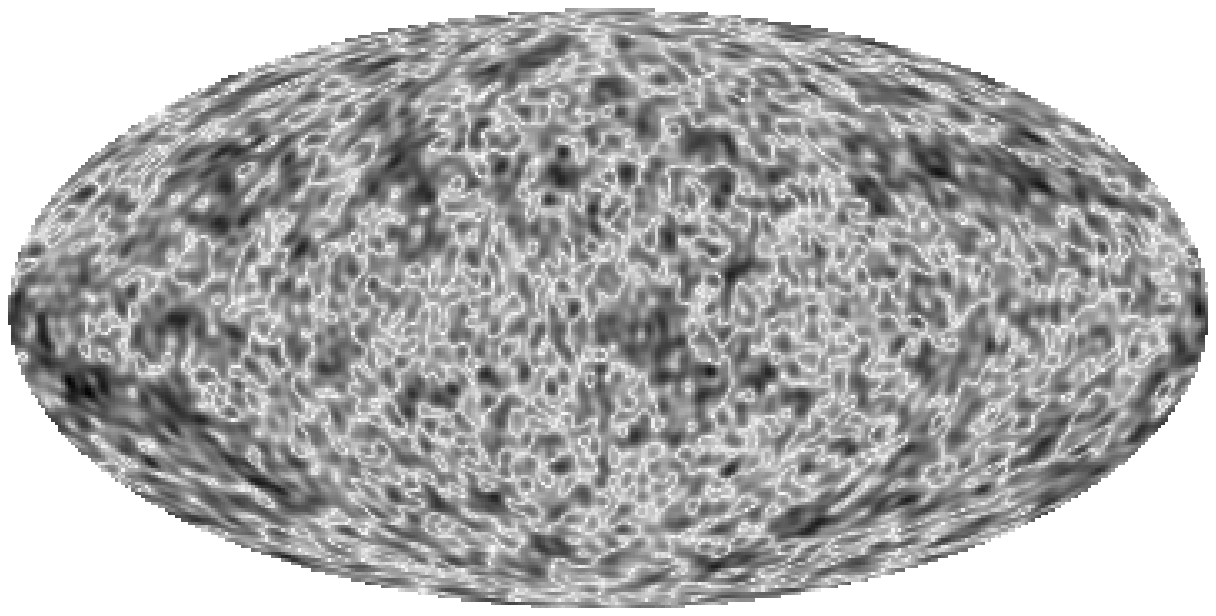}
\end{minipage}
\vspace*{-10pt}
\end{center}
\caption{\label{Fig:Picard_gen_ran_m30_q00_l65_h65_cusp}
The CMB anisotropy $\delta T$ for the Picard model for
$\Omega_{\hbox{\scriptsize mat}} = 0.3$ and
$\Omega_\Lambda = 0.65$ with a cut-off $k_c=140$
using the cusp forms only.
The observer is located in the upper half-space at $\vec x = (0.2, 0.1, 1.6)$.
}
\end{figure}
\begin{figure}[b]
\begin{center}
\hspace*{-150pt}\begin{minipage}{7cm}
\includegraphics[width=7.0cm,angle=270]{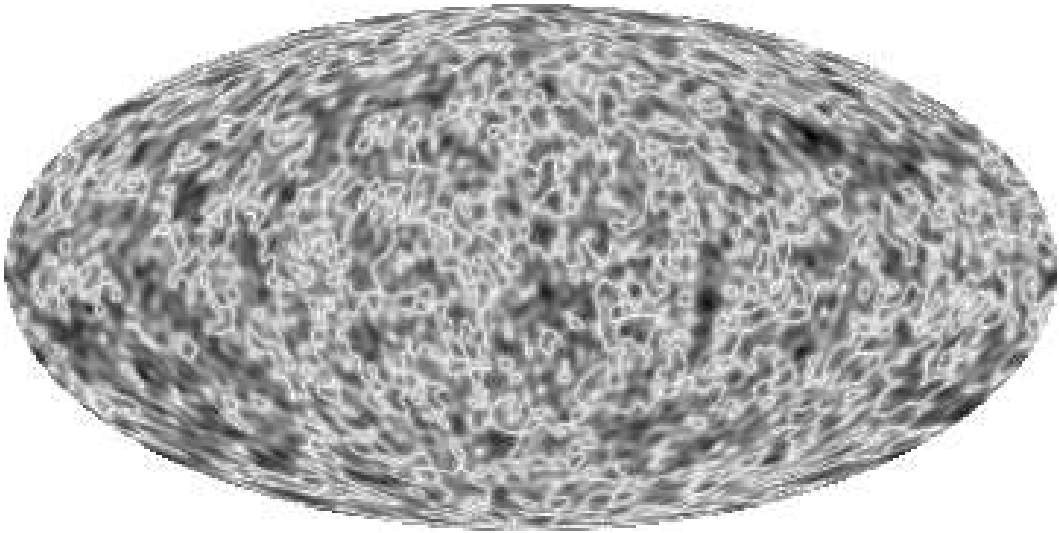}
\end{minipage}
\vspace*{-10pt}
\end{center}
\caption{\label{Fig:Picard_gen_ran_m35_q00_l60_h65_cusp}
The CMB anisotropy $\delta T$ for the Picard model for
$\Omega_{\hbox{\scriptsize mat}} = 0.35$ and
$\Omega_\Lambda = 0.60$ with a cut-off $k_c=140$
using the cusp forms only.
The observer is located in the upper half-space at $\vec x = (0.2, 0.1, 1.6)$.
}
\end{figure}
We use the Maa\3 cusp forms computed along the lines
described in \cite{Then_2003}.
(See \cite{Then_2003,Hejhal_1999} for further references concerning
the computation of the Maa\3 wave forms.
For earlier computations see \cite{Steil_1999}.)


\begin{figure}[htb]
\begin{center}
\hspace*{-160pt}\begin{minipage}{12cm}
\includegraphics[width=9.0cm]{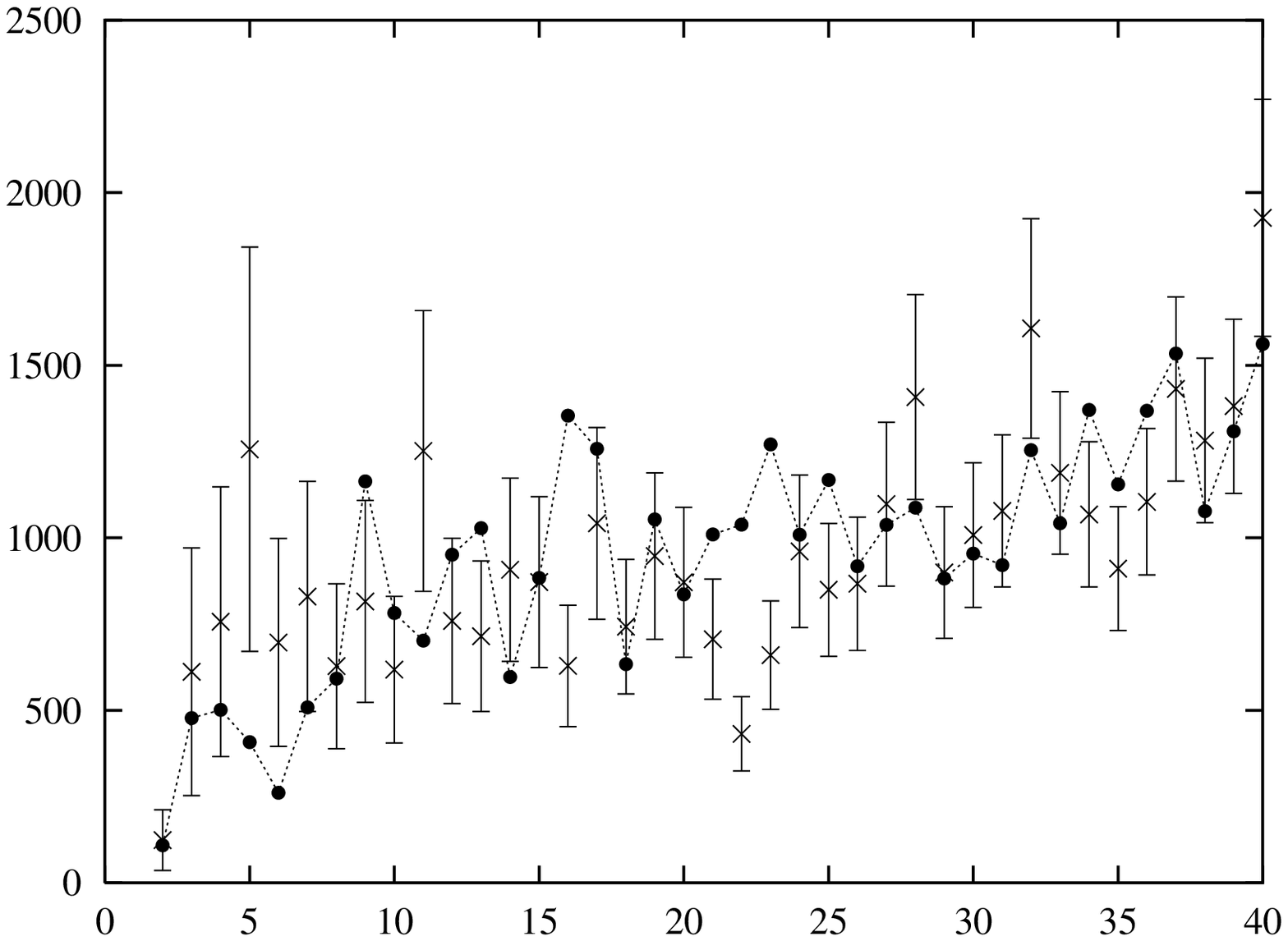}
\put(-1,4){$l$}
\put(-220,150){a)}
\put(-260,155){$\delta T_l^2$}
\put(-264,125){$[\mu \hbox{K}^2]$}
\hspace*{-10pt}\includegraphics[width=9.0cm]{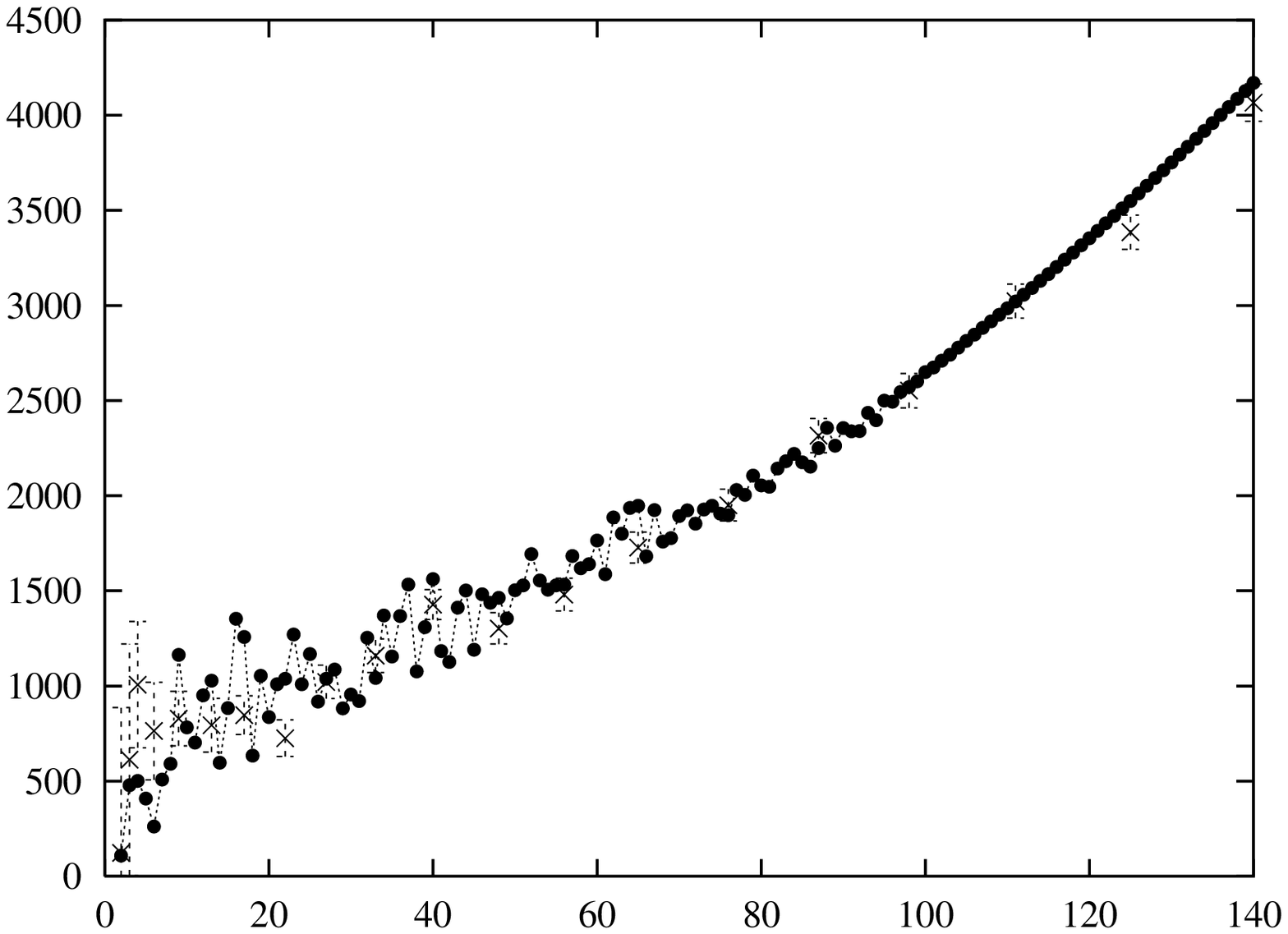}
\put(-220,150){b)}
\put(-1,4){$l$}
\end{minipage}
\vspace*{-10pt}
\end{center}
\caption{\label{Fig:Picard_m30_q00_l65_Cl_cusp}
The angular power spectrum $\delta T_l^2 = l(l+1) C_l/2\pi$ is
shown as dots for the sky map displayed in figure
\protect\ref{Fig:Picard_gen_ran_m30_q00_l65_h65_cusp}.
The cruxes with error bars are the first year WMAP data.
}
\end{figure}

\begin{figure}[htb]
\begin{center}
\hspace*{-160pt}\begin{minipage}{12cm}
\includegraphics[width=9.0cm]{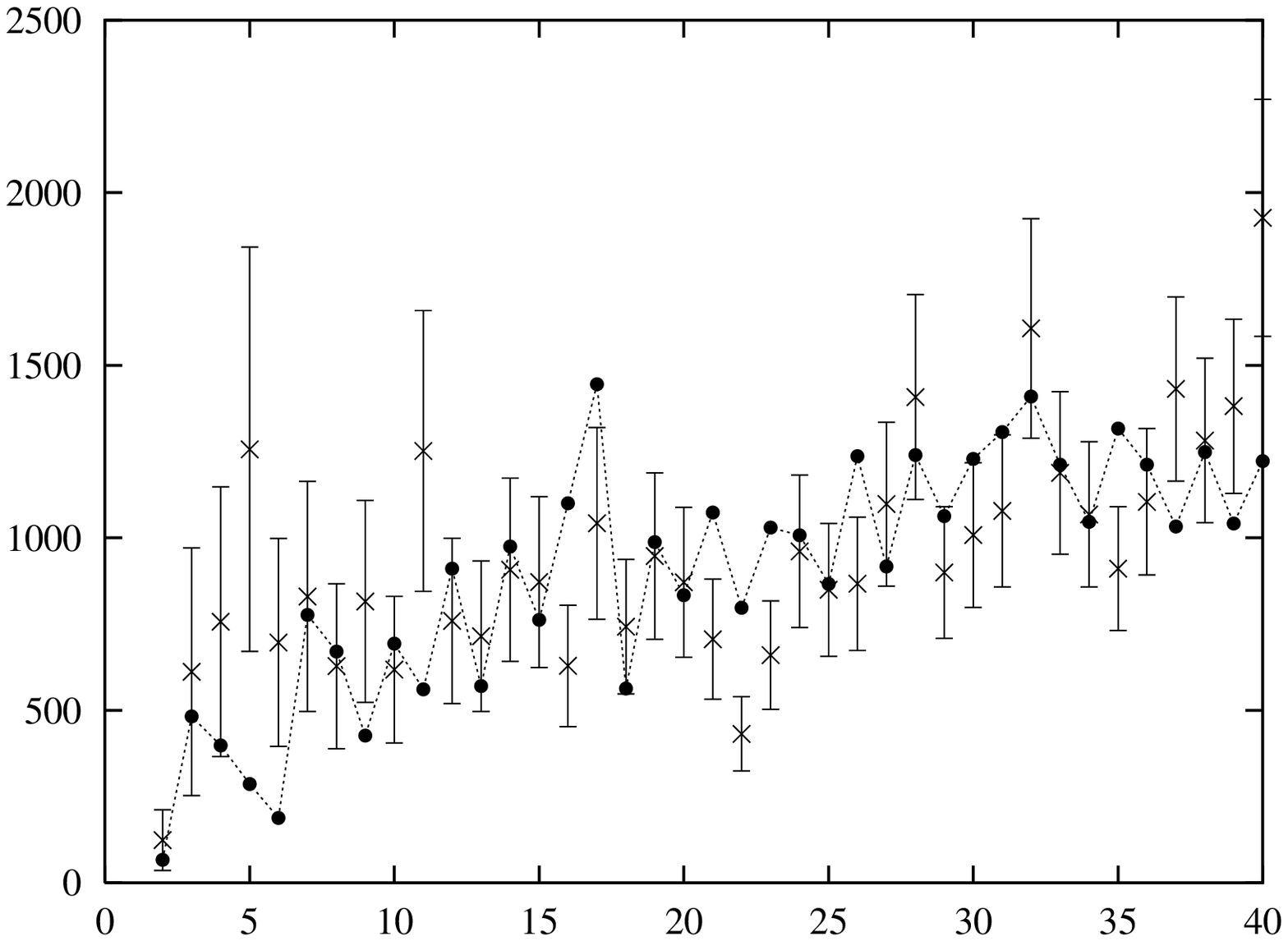}
\put(-1,4){$l$}
\put(-220,150){a)}
\put(-260,155){$\delta T_l^2$}
\put(-264,125){$[\mu \hbox{K}^2]$}
\hspace*{-10pt}\includegraphics[width=9.0cm]{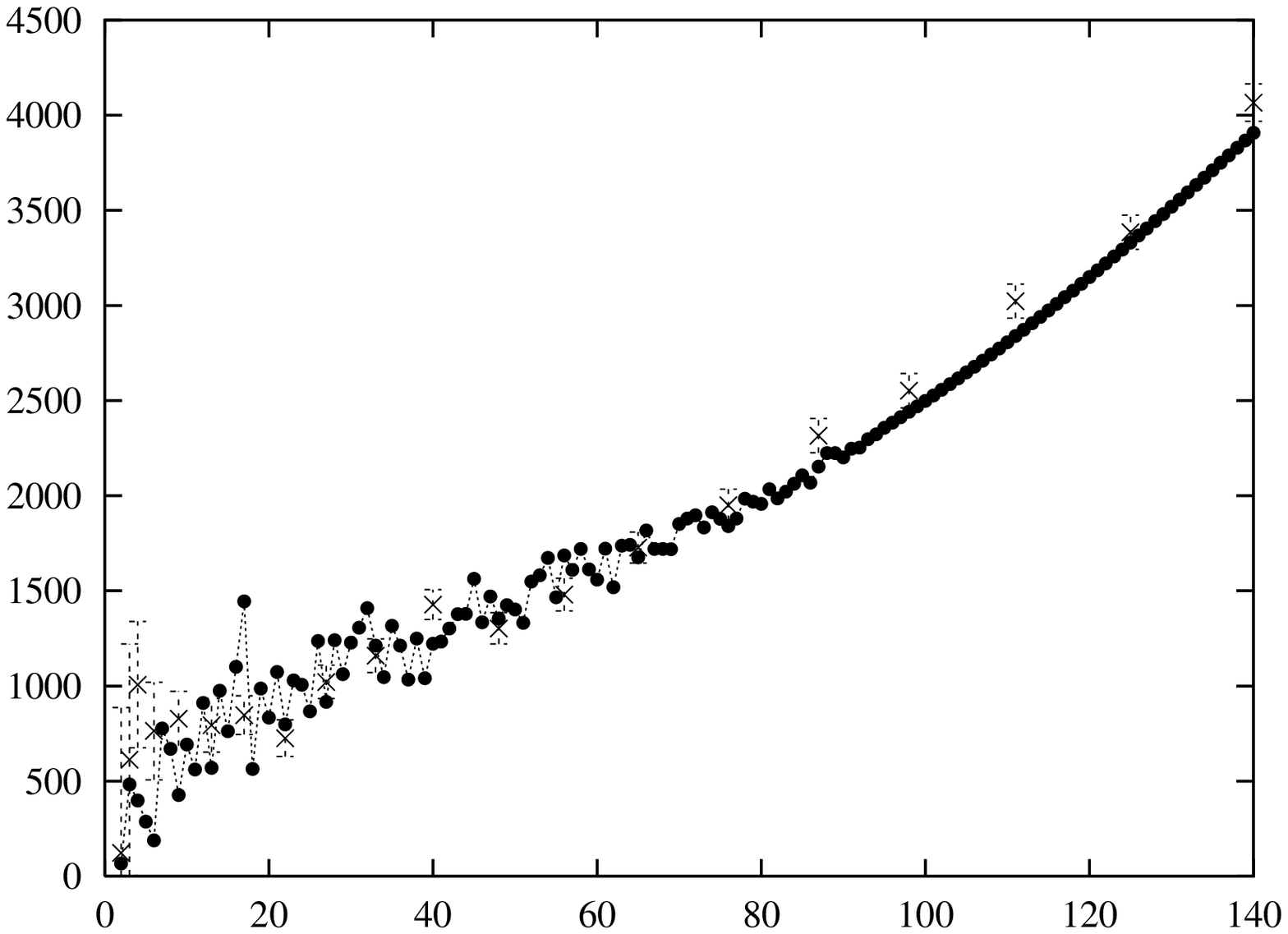}
\put(-220,150){b)}
\put(-1,4){$l$}
\end{minipage}
\vspace*{-10pt}
\end{center}
\caption{\label{Fig:Picard_m35_q00_l60_Cl_cusp}
The same as in figure \protect\ref{Fig:Picard_m30_q00_l65_Cl_cusp}
but for the sky map displayed in figure
\protect\ref{Fig:Picard_gen_ran_m35_q00_l60_h65_cusp}.
}
\end{figure}

\begin{figure}[htb]
\begin{center}
\hspace*{-10pt}\begin{minipage}{9cm}
\includegraphics[width=9.0cm]{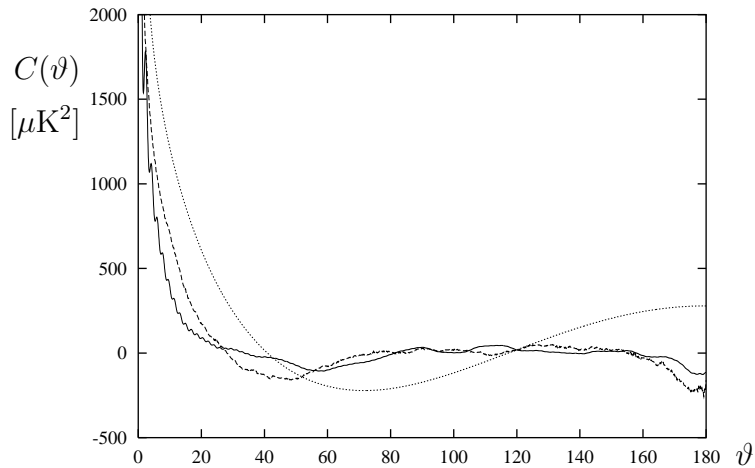}
\put(-1,4){$\vartheta$}
\put(-274,150){$C(\vartheta)$}
\put(-276,130){$[\mu \hbox{K}^2]$}
\end{minipage}
\vspace*{-10pt}
\end{center}
\caption{\label{Fig:Picard_gen_ran_m30_q00_l65_h65_C_theta}
The temperature correlation function $C(\vartheta)$
is shown as a full curve for the sky map displayed in figure
\protect\ref{Fig:Picard_gen_ran_m30_q00_l65_h65_cusp},
i.\,e.\ with $\Omega_{\hbox{\scriptsize mat}} = 0.3$ and
$\Omega_\Lambda = 0.65$.
The corresponding WMAP curve is shown as a dashed curve.
The dotted curve represents the best $\Lambda$CDM model
described in \cite{Bennett_et_al_2003}.
}
\end{figure}

\begin{figure}[htb]
\begin{center}
\hspace*{-10pt}\begin{minipage}{9cm}
\includegraphics[width=9.0cm]{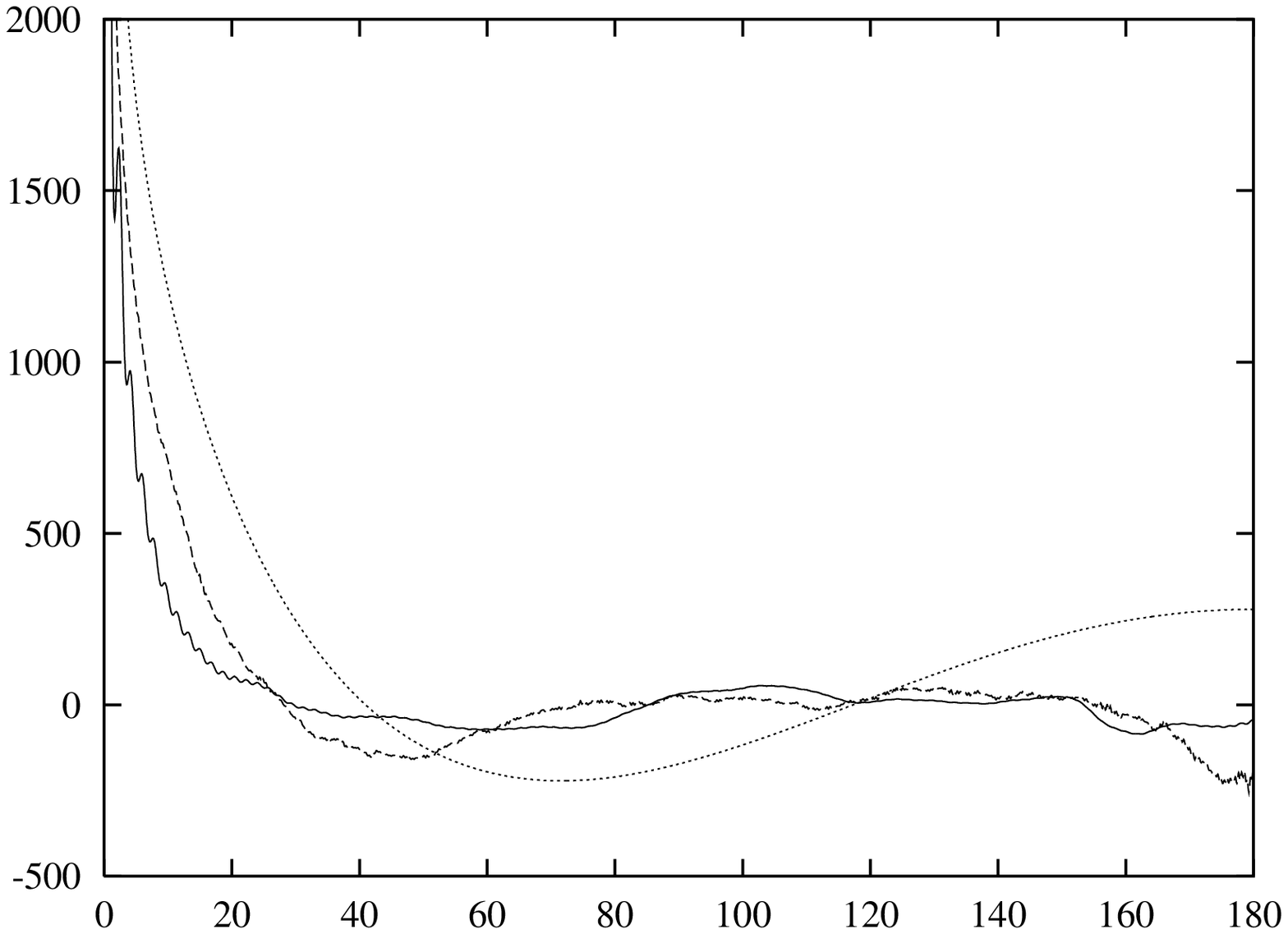}
\put(-1,4){$\vartheta$}
\put(-274,150){$C(\vartheta)$}
\put(-276,130){$[\mu \hbox{K}^2]$}
\end{minipage}
\vspace*{-10pt}
\end{center}
\caption{\label{Fig:Picard_gen_ran_m35_q00_l60_h65_C_theta}
The temperature correlation function $C(\vartheta)$
is shown as a full curve for the sky map displayed in figure
\protect\ref{Fig:Picard_gen_ran_m35_q00_l60_h65_cusp},
i.\,e.\ with $\Omega_{\hbox{\scriptsize mat}} = 0.35$ and
$\Omega_\Lambda = 0.6$.
The corresponding WMAP curve is shown as a dashed curve.
The dotted curve represents the best $\Lambda$CDM model
described in \cite{Bennett_et_al_2003}.
}
\end{figure}

In figures \ref{Fig:Picard_gen_ran_m30_q00_l65_h65_cusp} and
\ref{Fig:Picard_gen_ran_m35_q00_l60_h65_cusp} we present two
CMB simulations having $\Omega_{\hbox{\scriptsize tot}} = 0.95$,
the first having $\Omega_{\hbox{\scriptsize mat}} = 0.30$
and the second $\Omega_{\hbox{\scriptsize mat}} = 0.35$.
Here only the cusp forms up to $k_c=140$ are used.
The observer is located in the upper half-space at $\vec x = (0.2, 0.1, 1.6)$.
As in the case of the Sokolov-Starobinskii model
no suppression of anisotropy is observed towards the horn
which lies at the equator at one fourth from the left,
as in the previous sky maps.
The figures \ref{Fig:Picard_m30_q00_l65_Cl_cusp} and
\ref{Fig:Picard_m35_q00_l60_Cl_cusp} show the corresponding angular
power spectra $\delta T_l^2$.
These spectra have a very small quadrupole moment as it is observed by WMAP.
The next few moments are also well suppressed.
The $k$-summation in (\ref{Eq:Sachs_Wolfe_tight_coupling}) is
carried out for $k<140$ using the cusp forms.
The contribution of modes with $k>140$ is approximated
assuming statistical isotropy as in figure \ref{Fig:SoSta_Cl}
using the density of modes as it is given by Weyl's law
for the cusp forms \cite{Matthies_1995}.
The figures \ref{Fig:Picard_gen_ran_m30_q00_l65_h65_C_theta} and
\ref{Fig:Picard_gen_ran_m35_q00_l60_h65_C_theta}
demonstrate that both models describe the 
temperature correlation function $C(\vartheta)$ much better than
the concordance model.
Both models display a very small correlation at large scales,
$\vartheta \gtrsim 60^\circ$, as observed by WMAP.
In addition, the two models agree even at smaller scales,
$\vartheta \simeq 10^\circ$, with the observations much better than the
concordance model.
This is due to the finite volume of the Picard cell as can be seen
in comparison with figure \ref{Fig:SoSta_Merge3_C_theta} for
the Sokolov-Starobinskii cell which has infinite volume.
Figure \ref{Fig:SoSta_Merge3_C_theta} shows that the theoretical curve
for the Sokolov-Starobinskii model oscillates around the curve of
the concordance model.

\begin{figure}[htb]
\begin{center}
\hspace*{-150pt}\begin{minipage}{7cm}
\includegraphics[width=7.0cm,angle=270]{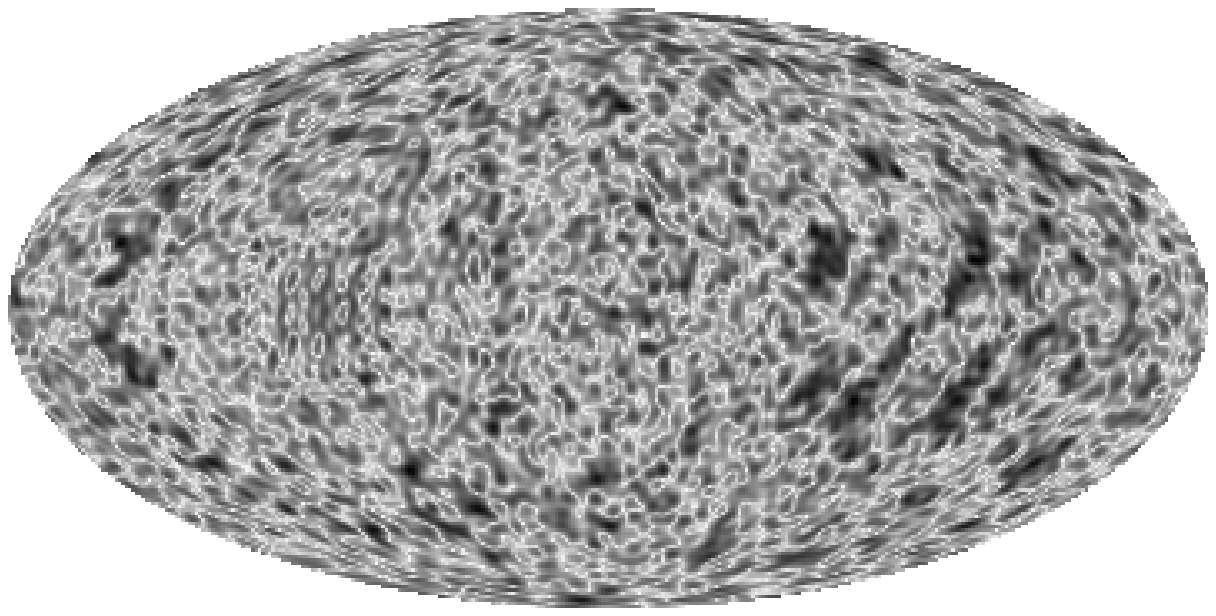}
\end{minipage}
\vspace*{-10pt}
\end{center}
\caption{\label{Fig:Picard_gen_Spitze_ran_m30_q00_l65_h65_cusp}
The CMB anisotropy $\delta T$ for the Picard model for
$\Omega_{\hbox{\scriptsize mat}} = 0.3$ and
$\Omega_\Lambda = 0.65$ with a cut-off $k_c=140$ using the cusp forms.
The observer is located in the upper half-space at $\vec x = (0.2, 0.1, 5.0)$.
}
\end{figure}

\begin{figure}[htb]
\begin{center}
\hspace*{-10pt}\begin{minipage}{9cm}
\includegraphics[width=9.0cm]{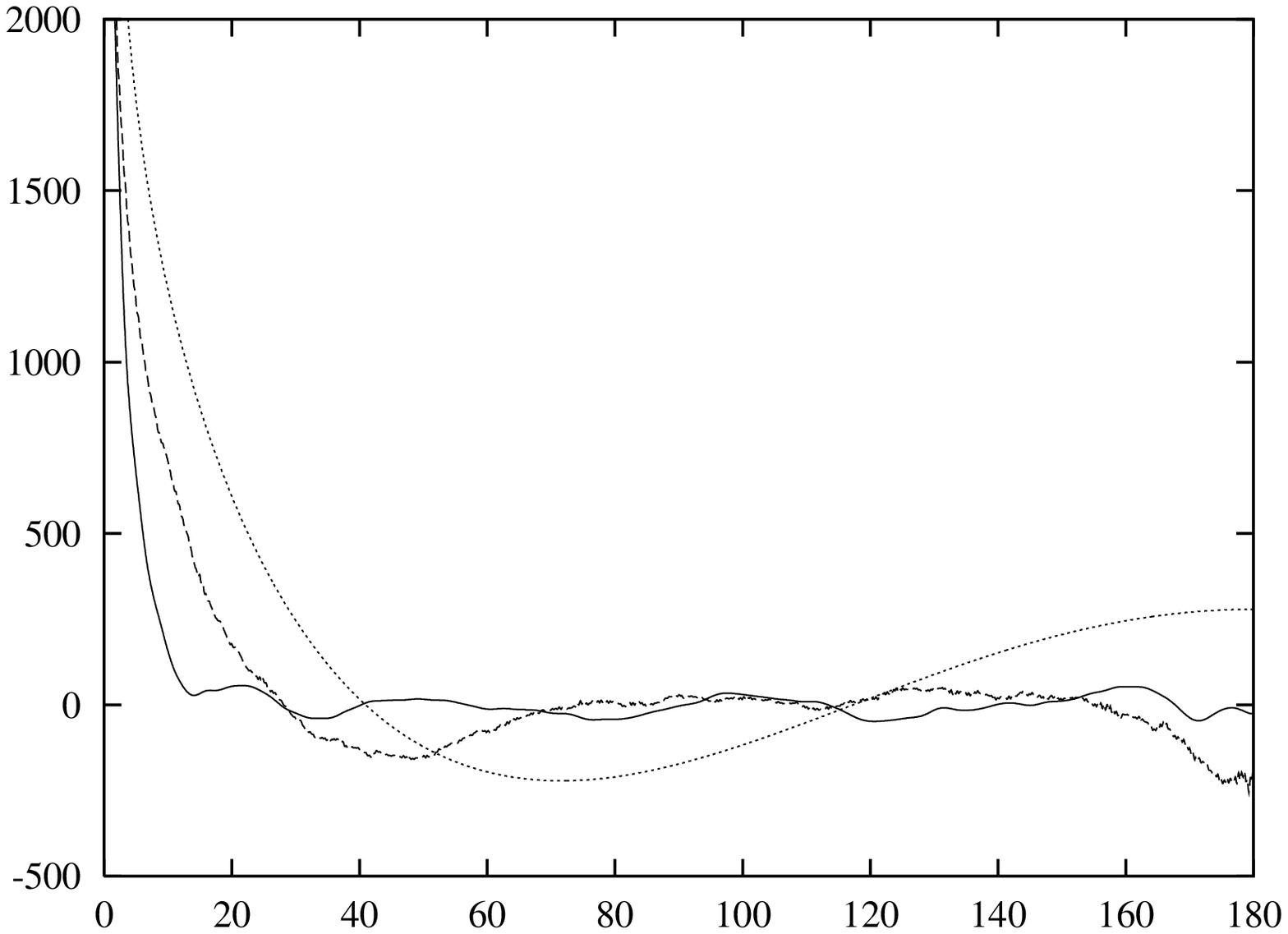}
\put(-1,4){$\vartheta$}
\put(-274,150){$C(\vartheta)$}
\put(-276,130){$[\mu \hbox{K}^2]$}
\end{minipage}
\vspace*{-10pt}
\end{center}
\caption{\label{Fig:Picard_gen_Spitze_ran_m30_q00_l65_h65_C_Theta_cusp}
The temperature correlation function $C(\vartheta)$
is shown as a full curve for the sky map displayed in figure
\protect\ref{Fig:Picard_gen_Spitze_ran_m30_q00_l65_h65_cusp},
i.\,e.\ with the observer sitting ``high'' in the horn and for
$\Omega_{\hbox{\scriptsize mat}} = 0.30$ and $\Omega_\Lambda = 0.65$.
The corresponding WMAP curve is shown as a dashed curve.
The dotted curve represents the best $\Lambda$CDM model
described in \cite{Bennett_et_al_2003}.
}
\end{figure}

In figure \ref{Fig:Picard_gen_Spitze_ran_m30_q00_l65_h65_cusp}
the CMB anisotropy $\delta T$ is shown for an observer who is
much higher in the horn at $\vec x = (0.2, 0.1, 5.0)$.
This is indeed very ``high'' in the horn as can be seen by comparing the volume
$V_\uparrow$ of the fundamental cell (\ref{Picard_Fundamental_Cell})
``above'' the observer in the direction of the horn,
i.\,e.\ the volume with $x_3>x_3^{\hbox{\scriptsize obs}}$, with the volume
$V_\downarrow = \hbox{vol}(\Gamma\backslash \HS) - V_\uparrow$ ``below''.
Due to the hyperbolic volume element $d\mu$, the volume above the observer is
$V_\uparrow = \frac 12 \int_{x_3^{\hbox{\scriptsize obs}}}^\infty d x_3 / x_3^3
= \frac 14 (x_3^{\hbox{\scriptsize obs}})^{-2}$,
$x_3^{\hbox{\scriptsize obs}}>1$.
In the previous simulations for the Picard model,
the observer was located at $x_3^{\hbox{\scriptsize obs}}=1.6$
for which one obtains
$V_\downarrow \simeq 0.207$ and $V_\uparrow \simeq 0.0976$, i.\,e.\ roughly
one third of the total volume lies in the direction of the horn.
For the observer of figure \ref{Fig:Picard_gen_Spitze_ran_m30_q00_l65_h65_cusp}
high in the horn with $x_3^{\hbox{\scriptsize obs}}=5$ one obtains
$V_\downarrow \simeq 0.295$ and $V_\uparrow = 0.01$.
Thus this observer point is very unprobable but we want to
stress here that even such an extreme position does not betray the
horn topology in the CMB anisotropy.
The figure \ref{Fig:Picard_gen_Spitze_ran_m30_q00_l65_h65_C_Theta_cusp}
displays the corresponding temperature correlation function $C(\vartheta)$.
Again there is very low correlation for $\vartheta \gtrsim 20^\circ$.
In this case the observed anticorrelation for $\vartheta \gtrsim 160^\circ$
is not well reproduced.
But nevertheless the overall agreement with WMAP data is much better
than for the concordance model.


\begin{figure}[htb]
\begin{center}
\hspace*{-150pt}\begin{minipage}{7cm}
\includegraphics[width=7.0cm,angle=270]{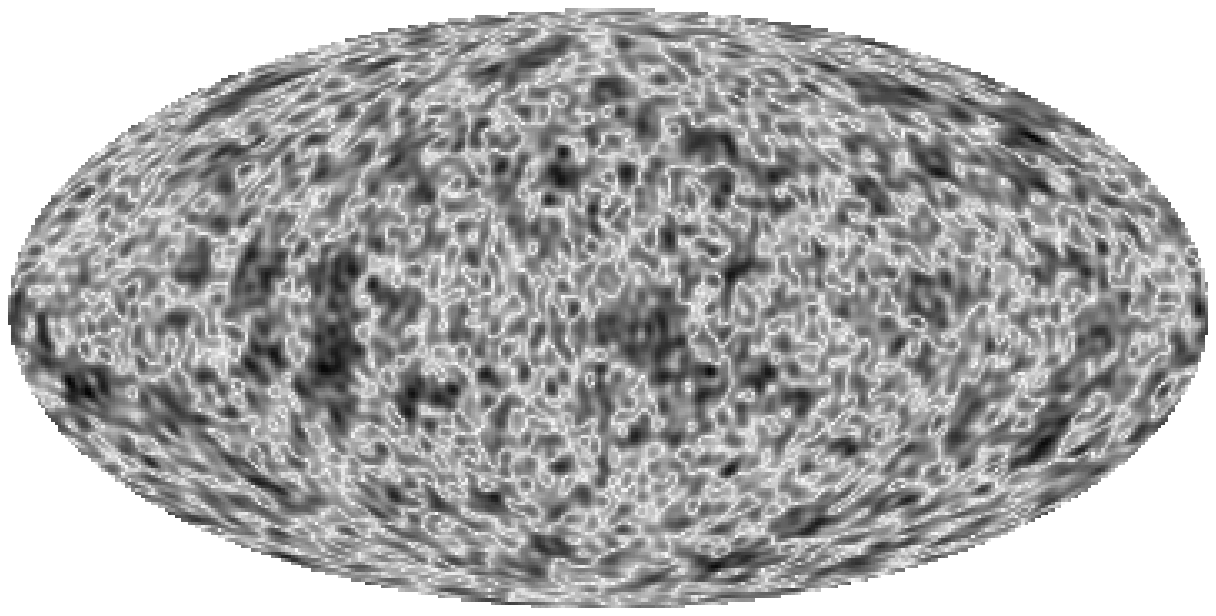}
\end{minipage}
\vspace*{-10pt}
\end{center}
\caption{\label{Fig:Picard_gen_ran_m30_q00_l65_h65_Merge}
The CMB anisotropy $\delta T$ for the Picard model for
$\Omega_{\hbox{\scriptsize mat}} = 0.3$ and
$\Omega_\Lambda = 0.65$ with a cut-off $k_c=140$
for a superposition of cusp forms and Eisenstein series
($\alpha = \frac 15$).
The observer is located in the upper half-space at $\vec x = (0.2, 0.1, 1.6)$.
}
\end{figure}

\begin{figure}[htb]
\begin{center}
\hspace*{-10pt}\begin{minipage}{9cm}
\includegraphics[width=9.0cm]{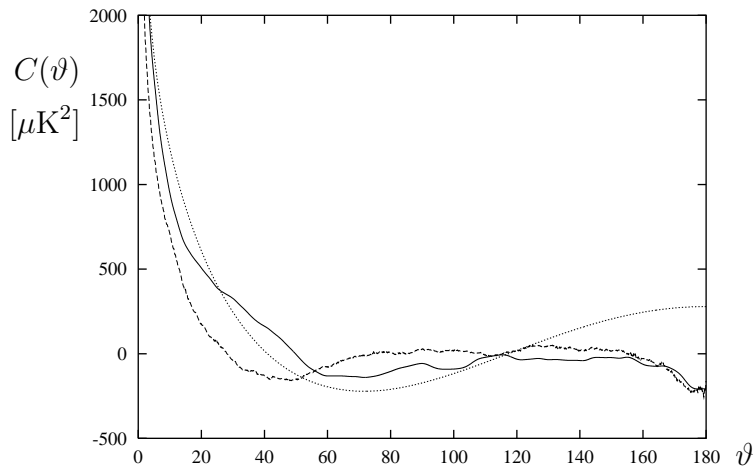}
\put(-1,4){$\vartheta$}
\put(-274,150){$C(\vartheta)$}
\put(-276,130){$[\mu \hbox{K}^2]$}
\end{minipage}
\vspace*{-10pt}
\end{center}
\caption{\label{Fig:Picard_gen_ran_m30_q00_l65_h65_Merge_C_Theta}
The temperature correlation function $C(\vartheta)$
is shown as a full curve for the sky map displayed in figure
\protect\ref{Fig:Picard_gen_ran_m30_q00_l65_h65_Merge}
using cusp forms as well as the Eisenstein series
($\Omega_{\hbox{\scriptsize mat}} = 0.30$ and
$\Omega_\Lambda = 0.65$).
The corresponding WMAP curve is shown as a dashed curve.
The dotted curve represents the best $\Lambda$CDM model
described in \cite{Bennett_et_al_2003}.
}
\end{figure}

Now let us consider a superposition of the contributions
of the cusp forms and the Eisenstein series.
The following model has $\Omega_{\hbox{\scriptsize mat}} = 0.3$ and
$\Omega_\Lambda = 0.65$.
In figure \ref{Fig:Picard_gen_ran_m30_q00_l65_h65_Merge}
the CMB anisotropy $\delta T$ is shown where the Eisenstein series
is weighted by a factor $\alpha=\frac 15$ relative to the contribution
of the cusp forms.
This factor depends on the chosen discretization of the $k$-integration
as discussed above.
Using a discretization finer than $N=16$,
i.\,e.\ 16 evaluation points per unit interval,
would allow a larger factor.
Since the summation of the cusp forms is not arbitrary,
there are values of $N$ for which this factor can be one.
The corresponding temperature correlation function $C(\vartheta)$
shown in figure \ref{Fig:Picard_gen_ran_m30_q00_l65_h65_Merge_C_Theta}
displays a very small correlation for $\vartheta \gtrsim 60^\circ$
and the observed anticorrelation for $\vartheta \gtrsim 160^\circ$ is matched
very well.


\section{Summary}

A large part of this paper was devoted to the question of the
existence of flat spots in the CMB sky maps for universes
with a horned topology.
By the example of two such universes, the Sokolov-Starobinskii model
and the Picard model, we showed that the infinitely long horn does
not lead to flat spots, i.\,e.\ to a suppression of the CMB fluctuations
in the horn, if the wavenumber cut-off $k_c$ is chosen sufficiently large.
The flat spots reported earlier \cite{Levin_Barrow_Bunn_Silk_1997,Levin_2002}
are thus seen as the result of taking not enough modes into account
in the expansion of the metric perturbation.
We conclude that the CMB sky maps do not reveal a signature of
the horned topology in form of flat spots.

Another main point of this paper was to show
that a universe with a horned topology, but with a finite volume,
such as the Picard model, can explain the loss of power at large
angular scales in the CMB anisotropy, as observed by COBE and WMAP.
However, if the volume is infinite, as in the Sokolov-Starobinskii model,
we have demonstrated that the low quadrupole moment is not reproduced.
There is, however, some controversy of
how serious one has to take this low value.
The small values of the first few multipole moments $C_l$
have, on the one hand, not been taken so serious
to require some new physics, but rather have been considered
as a manifestation of cosmic variance or an unsufficient
consideration of Galactic emission \cite{Efstathiou_2003}
or caused by the local supercluster \cite{Abramo_Sodre_2003}.
From this point of view the Sokolov-Starobinskii model is a viable model.
On the other hand, the suppression has been taken as a hint to new physics
such as new information on the inflation potentials
\cite{Cline_Crotty_Lesgourgues_2003,%
Contaldi_Peloso_Kofman_Linde_2003,Feng_Zhang_2003}.
However, these potentials have to be fine-tuned such
that the arising power spectra are suppressed around the present day
cosmological horizon \cite{Contaldi_Peloso_Kofman_Linde_2003}.
Such a fine-tuning is not necessary in the case of a non-trivial
topology in a universe with negative spatial curvature
due to the Mostow rigidity theorem \cite{Mostow_1973,Prasad_1973}.
Thus if the curvature scale is fixed by the densities $\Omega_x$,
all side-lengths of the fundamental cell ${\cal F}$ are determined,
and in turn the comoving wavenumber $k$ at which the spectrum
is suppressed, not because of the initial power spectrum
but simply because of the absence of modes below the lowest
wavenumber $k_1$.

In models with a non-trivial topology,
certain points at the surface of last scattering can be identical
due to the periodicity condition.
These matching points are located on pairs of circles with the
same radius.
Along two such circles the temperature fluctuations $\delta T$
produced by the naive Sachs-Wolfe effect are the same,
which is called the circles-in-the-sky signature
\cite{Cornish_Spergel_Starkman_1998b}.
In \cite{Cornish_Spergel_Starkman_Komatsu_2003} a search for
such circles in the WMAP data was carried out for nearly
back-to-back circles, i.\,e.\ for circles whose centers
have a distance greater than 170$^\circ$ and whose radii
are  greater than 25$^\circ$ on the sky.
They found no signature and rule out all topologies having
such circles.
However, it should be kept in mind that only the naive Sachs-Wolfe
contribution leads to identical temperature fluctuations.
The integrated Sachs-Wolfe contribution arises on the photon path
to the observer, which is not identified for the observer and
the ``copy'' of the observer.
The same is valid for the Doppler contribution, since the
observer and its copy see another projection of the velocity, in general.
Furthermore, there are a lot of other secondary contributions
to the temperature fluctuations.
Nevertheless, it is claimed in \cite{Cornish_Spergel_Starkman_Komatsu_2003}
that the naive Sachs-Wolfe contribution is strong enough for the
identification of circles-in-the-sky.
The Picard model is not ruled out by this work,
since there are no nearly back-to-back circles.
For the model with 
$\Omega_{\hbox{\scriptsize mat}} = 0.3$ and $\Omega_\Lambda = 0.65$
with $x_3=1.6$,
we obtain 40 pairs, where the largest distance of the centers is
at 145$^\circ$ which is not covered by the study in
\cite{Cornish_Spergel_Starkman_Komatsu_2003}.
The model with
$\Omega_{\hbox{\scriptsize mat}} = 0.35$ and $\Omega_\Lambda = 0.6$
has only 32 circle pairs.
The number of paired circles increases if the observer is posited
higher up in the horn, i.\,e.\ at larger values of $x_3$.
For the extreme position at $x_3=5$ one obtains
275 circle pairs with distances up to 168.6$^\circ$
for $\Omega_{\hbox{\scriptsize mat}} = 0.3$ and $\Omega_\Lambda = 0.65$.
These circles could have been detected in
\cite{Cornish_Spergel_Starkman_Komatsu_2003}.
However, for more generic observers, which are not sitting extremely
high in the horn, the separation of the circles is too small
such that the model cannot yet be ruled out.

In conclusion, we would like to emphasize again
that the Picard model studied in this paper is in nice agreement
with the observed suppression of power on large scales in
the angular power spectrum of the CMB
(see figures \ref{Fig:Picard_gen_ran_m30_q00_l65_h65_C_theta},
\ref{Fig:Picard_gen_ran_m35_q00_l60_h65_C_theta},
\ref{Fig:Picard_gen_Spitze_ran_m30_q00_l65_h65_C_Theta_cusp} and
\ref{Fig:Picard_gen_ran_m30_q00_l65_h65_Merge_C_Theta}).
This is in contrast to the concordance model
which does not reproduce the experimentally observed suppression
at $\vartheta \gtrsim 60^\circ$ and the observed correlation hole
at $\vartheta \gtrsim 160^\circ$.
If future observations will confirm the WMAP data but with smaller errors,
this can be interpreted as a clear hint to a non-trivial topology
of our Universe having negative spatial curvature and a finite volume.


\section*{Acknowledgment}

Financial support by the Deutsche Forschungsgemeinschaft (DFG)
under contract No Ste 241/16-1 and the
EC Research Training Network HPRN-CT-2000-00103 is gratefully acknowledged.

\section*{References}

\bibliography{../../bib_chaos,../../bib_astro}
\bibliographystyle{h-physrev3}

\end{document}